\documentclass[%
 reprint,
nofootinbib,
 amsmath,amssymb,
 aps,
]{revtex4-2}

\usepackage{graphicx}
\usepackage{dcolumn}
\usepackage{bm}
\usepackage{hyperref}
\hypersetup{
    colorlinks=true,
    linkcolor=[rgb]{0 0 0.5},
    urlcolor=cyan,
    citecolor= [rgb]{0 0 0.5}
}

\usepackage{soul} 
\usepackage{xcolor} 

\newcommand{\comment}[1]{}

\DeclareMathOperator{\Tr}{{\rm Tr}}

\newcommand{\bs}{\boldsymbol}

\newcommand{\BEA}{\begin{eqnarray}}
\newcommand{\EEA}{\end{eqnarray}}

\newcommand{\la}{\left\langle}
\newcommand{\ra}{\right\rangle}
\newcommand{\tr}[1]{#1^\intercal}

\makeatletter
\@ifundefined{ol}
  {\newcommand{\ol}[1]{\overline{#1}}}
  {\renewcommand{\ol}[1]{\overline{#1}}}
\makeatother

\renewcommand{\d}{{\rm d}}
\newcommand{\ee}{{\rm e}}
\newcommand{\ii}{{\rm i}}
\newcommand{\pd}[2]{\frac{\partial #1}{\partial #2}}

\newcommand{\curl}{\text{curl\,}}
\newcommand{\sgn}{{\rm sgn}}

\renewcommand{\v}{{\bs v}}

\renewcommand{\o}{{\bs\Omega}}
\newcommand{\uo}{\,\underline{\Omega}\,}
\newcommand{\rI}{\,\underline{{\rm I }}\,}
\newcommand{\ez}{\bs{e}_z}
\newcommand{\bxi}{{\bs \xi}}

\newcommand{\pp}{{}_{\!\scriptscriptstyle\bs{\perp}\!}}

\newcommand{\sfR}{\mathsf{R}}
\newcommand{\R}{{\bm R}}
\newcommand{\V}{{\bm V}}
\newcommand{\A}{{\bm A}}

\newcommand{\rS}{{\rm S}}
\newcommand{\rC}{{\rm C}}
\newcommand{\rK}{{\rm K}}
\newcommand{\rM}{{\rm M}}
\newcommand{\rN}{{\rm N}}
\newcommand{\rU}{{\rm U}}
\newcommand{\rH}{{\rm H}}
\newcommand{\rD}{{\rm D}}

\renewcommand{\P}{{\bm P}}

\renewcommand{\L}{{\bs L}}

\newcommand{\calL}{{\cal L}}
\newcommand{\calP}{{\cal P}}

\newcommand{\B}{{\bm B}}
\renewcommand{\r}{{\bm r}}

\newcommand{\eff}{{\rm eff}}
\newcommand{\Teff}{T^\eff}
\newcommand{\Oeff}{\Omega^\eff}

\renewcommand{\Re}{{\rm Re}}

\newcommand{\td}{\tilde}

\newcommand{\WK}{Wiener-Khinchin}

\newcommand{\TODO}[1]{[{\textbf{\hl{TODO}}}:  \textit{#1}]}

\definecolor{darkred}{rgb}{0.8,0,0}

\newcommand{\Ei}{\text{Ei}}

\begin{document}

\preprint{.}

\title{System-Bath Approach to Rotating Brownian Motion } 

\author{\vspace{-0.5ex}Ashot Matevosyan$^{1,2)}$ and Armen E. Allahverdyan$^{2)}$ }

\affiliation{$^{1)}$Max Planck Institute for the Physics of Complex Systems,\\
 N\"{o}thnitzerstra{\ss}e 38, Dresden, 01187, Germany\\
$^{2)}$Alikhanyan National Laboratory (Yerevan Physics Institute), \\ 
Alikhanian Brothers Street 2, Yerevan 375036, Armenia}





\begin{abstract}
Rotating equilibrated systems are widespread, but relatively little attention has been devoted to studying them from the first principles of statistical mechanics. We fill this gap by studying 
a Brownian particle coupled with a thermal bath made of rotating harmonic oscillators. We show that the Langevin equation that describes the dynamics of the Brownian particle contains (due to rotation) long-range correlated noise.
In contrast to the usual situation of (non-rotating) equilibration, the rotating Gibbs distribution is recovered only for a weak coupling with the bath. In the presence of a uniform magnetic field, the stationary state is not Gibbsian, even under weak coupling. In this context, we clarify the applicability of the Bohr-van Leeuwen theorem to classical systems in rotating equilibrium, as well as the concept of work done by a changing magnetic field. We show that the Brownian particle under a rotationally symmetric potential reaches a stationary state that behaves as an effective equilibrium, characterized by a free energy. As a result, no work can be extracted via cyclic processes that respect the rotation symmetry. However, if the external potential exhibits asymmetry, then work extraction via slow cyclic processes is possible. This is illustrated by a general scenario involving a slow rotation of a non-rotation-symmetric potential. We study sedimentation equilibrium and show that centrifugal instability is prevented by a finite friction. 

\end{abstract}

\maketitle


\section{Introduction}

\subsection{Gibbs distribution}

Thermodynamic systems are composed of microscopic constituents with well-defined quantum or classical dynamics. Statistical mechanics bridges the gap between macro and micro descriptions and accounts for phenomenological thermodynamics and non-equilibrium phenomena in physics \cite{terletski,landau_stat,reichl,de2013non,berdichevsky,risken,lp,tuckerman}. 
Problems in statistical mechanics typically consider a large system divided into two unequal parts: a smaller subsystem of interest, referred to as the ``system,'' and a larger quasi-equilibrium subsystem known as the ``thermostat,'' ``reservoir,'' or ``bath'' \cite{landau_stat}. The system interacts with the bath, exchanging energy, particles, linear momentum, and, as we will explore later, angular momentum \cite{matevosyan2024weak,lasting}. A fundamental concept in equilibrium statistical mechanics is the Gibbs distribution, which links the probability of a system in phase space with its Hamiltonian. Thermodynamic quantities are deduced as averages over the Gibbs distribution. 

The convergence of the system's dynamics toward the Gibbs distribution is less universal, as it depends on the details of the system and bath. The generalized, classical Langevin equation emerged recently as a sufficiently general paradigm for non-equilibrium dynamics and relaxation toward the Gibbs distribution \cite{tuckerman}. It is also deduced systematically from the system-bath approach \cite{tuckerman}. In particular, the projection operator formalism provides a powerful framework for deriving the generalized Langevin equation directly from the Hamiltonian of a closed system by examining the dynamics of slow degrees of freedom \cite{mori1965transport,zwanzig2001nonequilibrium, ayaz2022generalized}. The Caldeira-Leggett (CL) model can be viewed as a special case of this formalism: the bath is modeled as a collection of independent harmonic oscillators, each representing a mode of the thermal environment \cite{magalinskii1959dynamical,zwanzig1973nonlinear,caldeira1983quantum,tuckerman}. 
This model has a closed-form solution, reproduces the generalized Langevin equation for an initially equilibrium bath, and hence leads to the (non-rotating) Gibbs distribution for the system in the long-time limit. Oscillators in the CL model can represent phonons or photons, or (more generally) refer to the degrees of freedom of a weakly perturbed bath \cite{caldeira1983quantum,tuckerman}.  

The usual Gibbs distribution includes only energy and temperature. Here, we focus on a system that interacts with a rotating thermal bath. All interactions are assumed to be pairwise and rotation symmetric; that is, the total angular momentum is conserved for the overall system. As an additive conserving quantity, angular momentum affects the equilibrium statistical mechanics of the bath, since it now needs to be accounted for on an equal basis with energy \cite{landau_stat,jaynes}.
Hence, the rotating Gibbs density for the coordinates and velocities of the bath reads \cite{landau_stat,jaynes}:
\begin{equation}\label{general}
    \rho_B\propto \exp\left(-\beta \,H_{B}+\beta\,\bs\Omega \cdot  \L_{B} \right),
\end{equation}
where $H_{B}$, $\L_{B}$, and $\bs\Omega$ are (resp.) the Hamiltonian, angular momentum, and angular velocity of the bath. 
$H_{B}$ and $\L_{B}$ are functions of the positions and momenta of the bath. $H_{B}$ is rotation symmetric over these vectors. $\bs\Omega$ is a constant. 
$T=1/\beta$ ($k_{\rm B}=1$) is the temperature.

Eq.~\eqref{general} was proposed by Gibbs himself \cite{jaynes}; it is easily generalized to the microcanonical situation. Eq.~\eqref{general} can be deduced via the maximum entropy method and applied to an isolated macroscopic system with conserved angular momentum. It contains two additive conserved quantities. Both come from the assumed symmetries of the bath. For the bath that rotates with a constant angular speed, \eqref{general} can be derived by looking at thermalization in the rotating, non-inertial reference frame (assuming that all interactions are rotation symmetric) \cite{landau_stat,jaynes}. 
This derivation is based on the fact that the Hamiltonian becomes time-independent \footnote{In a non-inertial reference frame there is a constant term related to the rotation, i.e., the Hamiltonian is modified, but is not time-dependent \cite{landau_stat}.  } in the rotating frame, which allows the application of the thermalization argument \cite{jaynes}. Eq.~\eqref{general} is an example of generalized Gibbs ensembles, which have been intensively studied in the recent decade with a focus on integrable quantum systems \cite{emil}; see \cite{polkovnikov} for a review.

\subsection{Rotation and Rotating Equilibrium}
\label{exoroto}

Rotating motors play a crucial role in energy storage and transmission in molecular machines and nanodevices \cite{rotatory_nanomotors,venturi,splendid, marvellous,mugnaiRMP}. Interesting examples of rotating equilibrium motion are found in atomic clusters \cite{clusters} and in complex plasma \cite{bonitz-ott}. One category of rotating equilibrium systems includes classical Brownian charges, where a constant magnetic field causes the phonon bath to gain orbital momentum \cite{lasting}. Rotating quasi-equilibrium is also observed in self-gravitating star clusters \cite{laliena}.

There is an interesting application of rotating equilibration to ultracentrifuges, devices that rapidly rotate and are used to sediment complex fluids \cite{sedimentation_katchalsky,sedimentation1,sedimentation2}. Sedimentation is employed and studied both in stationary (sedimentation equilibrium) and non-stationary settings \cite{sedimentation1,sedimentation2}. Both these aspects can be phenomenologically described via the Lamm equation \cite{sedimentation2}, which is closely related to the phenomenological Langevin equation; see section \ref{pheno} and Ref.~\cite{matevosyan2024weak}. This equation converges to the rotating Gibbs distribution \eqref{general}. Hydrodynamic approaches to sedimentation were worked out in Refs.~\cite{rotating_brownian_prl, rotating_brownian_jsp, rotating_brownian_physicaa}. No consensus has been reached in these works about how to describe sedimentation efficiently. Hence, sedimentation (and centrifugation) is an applied field where microscopic approaches to rotating Brownian motion can be useful.  

In this context, one needs to distinguish between the Brownian motion under a rotating thermal bath (i.e., the bath performing regular rotation in addition to its random motions), and 
{\it rotational Brownian motion}, where a usual, non-rotating bath exerts a random torque on non-point-like Brownian particle\footnote{Debye laid down the foundations of the latter subject, which attracted much attention due to its molecular and atomic applications; see Refs.~\cite{rotation_relaxation_1, rotation_relaxation_2,rotation_relaxation_3} for recent results.}.

\subsection{Our results and the structure of the paper}

Here we aim to describe the Brownian motion for a particle coupled with a rotating bath. As opposed to previous phenomenological approaches \cite{bonitz-ott,kahlert2012magnetizing,matevosyan2024weak}, we take a more microscopic stance. We start from the bath of harmonic oscillators \cite{magalinskii1959dynamical,zwanzig1973nonlinear,caldeira1983quantum,tuckerman} (Caldeira-Leggett model), where the rotation is imposed in the initial state of the bath via the Gibbsian density \eqref{general}. The advantage of harmonic oscillators is that the bath dynamics is exactly solvable. 
As a result, Brownian particle dynamics are exactly like Langevin equation dynamics (exact means without further approximations). While the friction force of this Langevin equation is standard, e.g., it admits localization towards the usual Ohmic limit, the noise is long-range correlated due to rotation, even in the Ohmic limit. Both these aspects differ from the phenomenological description of Brownian motion due to a rotating bath proposed in Refs.~\cite{bonitz-ott,kahlert2012magnetizing,matevosyan2024weak}. 

The rotating Langevin equation is similar to the quantum Langevin equation, where (even for Ohmic friction) the noise is not white and shows long-range correlations due to quantum fluctuations; see, e.g., Ref. \cite{an}. Importantly, the long-time limit of the rotating Langevin equation reduces to the rotating Gibbsian state (i.e. the analog of \eqref{general}) of the Brownian particle only in the weak coupling limit, an aspect which is also similar to the quantum Langevin equation \cite{an}. Additional non-Gibbsian features for the stationary state exist whenever (for a rotating bath) an external homogeneous, time-independent magnetic field acts on the charged Brownian particle. We show that, despite the stationary state of a Brownian particle in a harmonic potential being non-Gibbsian \footnote{The non-Gibbsian state admits an effective Gibbsian description via effective potential and angular velocity; see sections \ref{sec-aomega} and \ref{sec-weak-coupling}. The effective description, however, does not by itself imply the existence of the free energy.}, it still admits a well-defined free energy: no work can be extracted by slowly and cyclically varying the system's external parameter. The existence of the free energy is, however, closely tied to the rotation symmetry of the potential acting on the Brownian particle: cyclic work extraction is possible if the external potential for the Brownian particle is non-rotation-symmetric.

Recall that when the rotation is absent, the classical Langevin equation always relaxes to the classical Gibbs density for the Brownian particle (provided only that the external potential is confining, as we shall always assume in this work) \cite{tuckerman,breuer2002theoryofopenquantum}. Likewise, without rotation, the external magnetic field does not enter into the coordinate-velocity stationary probability density for the classical Brownian particle. This is the known Bohr-van Leeuwen theorem, whose meaning was clarified recently \cite{lasting}; a brief summary of this theorem can be found at Appendix \ref{app-bvl}. 

The paper is structured as follows: In the next section, we define our microscopic model of the rotating thermal bath as an extension of the Caldeira-Leggett model. There, we also discuss stability features of the bath and deduce the Langevin equation for the Brownian motion due to a rotating bath. In section \ref{sec-aomega}, we discuss a solvable example of this Langevin equation. Section \ref{sec-ft} focuses on the stationary state of the Brownian particle interacting with the rotating bath.
Section \ref{sedi} applies this theory to sedimentation equilibrium and shows that for the present approach, the centrifugal instability is absent. The mechanism of this closely relates to a finite friction and the ensuing non-Gibbsianity of the stationary state. Section \ref{magno} looks at the charged particle subject to both the rotating thermal bath and external magnetic field. In section \ref{thermo}, we discuss the thermodynamics of the above stationary state, focusing on the isotropic (angular-momentum preserving) external potential for the Brownian particle. To make the presentation self-contained, section \ref{reminder} recalls the standard notion of free energy. Section \ref{freedom} discusses the work done by a time-dependent magnetic field. Section \ref{result} shows that the notion of  free energy still applies to the stationary state of the Brownian particle interacting with the rotating thermal bath and external magnetic field. The condition of isotropic potential is lifted in section \ref{sec-aniso}. A comparison of the present microscopic approach with phenomenological approaches is provided in section \ref{pheno}.  
We summarize in the last section.

\section{The Model} 
\label{sec-cl}

\begin{figure}[t]
\includegraphics[width=0.7\linewidth]{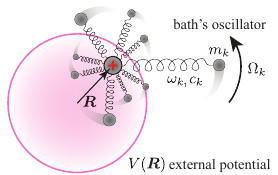}
\caption{Illustration of the system+bath. The system is a Brownian charge coupled to harmonic oscillators (thermal bath) with strength $c_k$. Each oscillator $k$ with frequency $\omega_k$ rotates with its own frequency $\Omega_k$; see \eqref{stability}. }
\label{fig0}
\end{figure}

\subsection{Equations of Motion}

Consider a classical particle with coordinates $\R = (X, Y, Z)$, unit charge, and unit mass interacting with a magnetic field. The particle is subject to an external, rotation symmetric potential $V(\R)$. The Lagrangian for the system is given by:
\BEA
\label{lagr_sys}
\calL_{S}=\frac{1}{2}\dot{\R}^2 - V(\R) + \A(\R)\dot{\R},
\EEA
where $\A(\R)$ is a vector potential generating a static, homogeneous magnetic field $\B$ with magnitude $b = |\B|$ along the $z$-axis (normal vector $\ez$). For $\A$ we take a version of the Coulomb gauge:
\BEA
\label{mag}
&&\B=\curl \A=\ez b,\\
&&\A(\R)
=\frac{1}{2}(-{bY}, {bX}, 0) = \frac{1}{2}\B\times\R.
\label{gauge-choice}
\EEA
The advantage of (\ref{gauge-choice}) is that the Euler-Lagrange equations of motion are invariant when coordinates and velocities are subject to rotations in the $(x,y)$-plane; see section \ref{freedom} for further details. 

The particle interacts with a bath consisting of $N$ harmonic oscillators (modes) characterized by coordinates $\r_k$, masses $m_k$, frequencies $\omega_k$, and coupling constants $c_k$. The potential energy of the particle-bath interaction is assumed to be non-negative and bilinear over the particle and bath coordinates (Caldeira-Leggett model); see Fig.~\ref{fig0}. Hence the bath+interaction Lagrangian reads \cite{magalinskii1959dynamical,breuer2002theoryofopenquantum,zwanzig1973nonlinear,caldeira1983quantum,tuckerman}
\BEA
\label{lagr_bath}
\calL_{B}={\sum}_{k=1}^N \left[\frac{m_k}{2}\dot{\r}_k^2 -
\frac{m_{k}\omega_{k}^{2}}{2} \left(\r_{k}-\frac{c_k\R}{m_k\omega_k^2}\right)^{2}\right],
\EEA
where the full Lagrangian is $\calL_{S}+\calL_{B}$. 
Solving the Euler-Lagrange equations of motion generated by $\calL_{S}+\calL_{B}$ for $\r_k$, we get: 
\BEA
    \r_k(t)&=&\r_k(0) \cos(\omega_k t) + \frac{\dot{\r}_{k}(0)}{\omega_k} \sin(\omega_k t)
    \\
    &&+\frac{c_k}{m_k\omega_k} \int_0^t \d t' \sin(\omega_k (t-t')) \R(t').
    \label{osa}
\EEA
Plugging \eqref{osa} into the Euler-Lagrange equations of motion for $\R$, we derive the Langevin equation for the charged particle in a magnetic field \cite{karmeshu1974}:
\BEA
\label{lang}
&&\ddot{\R}=b\,\ez\times\dot{\R}-\partial_\R V(\R)-\!\int_0^t\!\!\d u\,\zeta(t-u)\dot{\R}(u)+\bxi,\\
\label{kernel}
  &&  \zeta(t)=\!{\sum}_{k=1}^N \frac{c_k^2}{m_k \omega_k^2} \cos(\omega_k t), \\ 
  \label{noise}
    && \bxi(t) = \!{\sum}_{k=1}^N[\, c_k\overline{\r}_k(0) \cos(\omega_k t)
    \!+\! \frac{c_k \v_{k}(0)}{\omega_k} \sin(\omega_k t)],\,\,~~\\
&&\overline{\r}_k(0)\equiv\r_k(0)-\R(0)\frac{ c_k}{m_k \omega_k^2}, \qquad \v_k=\dot{\r}_k. 
\label{ido}
\EEA
The cumulative force from the bath is decomposed into non-local friction with a kernel $\zeta$ and noise $\bxi(t)$ that emerges due to the initially random state of the bath. Eq.~\eqref{lang} has the formal structure of the generalized Langevin equation, although the distribution of noise has not been specified yet. It will come with the initial state of the bath.

From Noether's theorem, or directly from the equations of motion generated by the full Lagrangian $\calL_{S}+\calL_{B}$, we find the following conserved quantity:
\BEA
    \Upsilon &=& X\dot{Y} - Y\dot{X} 
    + {\sum}_{k=1}^N m_k\left(x_k \dot y_k - y_k \dot x_k\right) 
    \\
    &&+ \frac{b}{2}\left(X^2 + Y^2\right).
    \label{momentum}
\EEA
Here, $\Upsilon$ represents the sum of the particle angular momentum $X\dot{Y} - Y\dot{X}$ along the $z$ axis [selected due to \eqref{mag}], the angular momenta of all bath oscillators, and an additional term $\frac{b}{2}\!\left(X^2 \!+\! Y^2\right)$ arising from the charged particle's interaction with the magnetic field. If the rotating Gibbs distribution \eqref{general} is written under an external magnetic field, $\Upsilon$ from \eqref{momentum} should appear instead of the angular momentum in \eqref{general}. Note that if no angular momentum is included in the Gibbs distribution (i.e. the term $\propto \Upsilon$ is absent), the external magnetic field does not appear in the stationary density for coordinates and velocities. This is the Bohr-van Leeuwen theorem; see Appendix \ref{app-bvl} and Ref.~\cite{lasting} for recent discussion of this theorem.

\subsection{The Initial State of the Bath}

\comment{
We assume that before $t=0$, the bath equilibrated under a fixed value of $\R$. The conditional equilibrium state of the bath is then described by the rotating Gibbs distribution [cf. \eqref{general}], which has the form of a conditional probability density 
\begin{align}
\label{initial_state}
&\rho\left(
\{ \r_k, \v_k\} \,|\, \R\,
\right) \propto 
\exp\Big(- \beta H_B\,+
\\
&\qquad\qquad+\beta\sum_{k=1}^N \Omega_k m_k\left(\ol{\r}_{k,x}\v_{k,y}-\ol{\r}_{k,y}\v_{k,x}\right) \Big),
\label{initial_state-l2}
\\
&H_B=\sum_{k=1}^N \left[\frac{m_k}{2}\dot{\r}_k^2 +
\frac{m_{k}\omega_{k}^{2}}{2} \ol{\r}_{k}^{2}\right].
\end{align}
Here, $H_B$ is the bath energy derived from \eqref{lagr_bath}. The conditional dependence on $\R$ arises from the relative coordinate $\ol{\r}$, as defined in \eqref{ido}. According to the angular momentum term in \eqref{initial_state}, each bath oscillator rotates around the $z$-axis passing through $\R$. 
Unlike in \eqref{general}, each oscillator has its own angular frequency $\Omega_k$; this is explained below.
Notably, both the energy $H_B$ and the angular momentum term in \eqref{initial_state}
are conserved quantities of the bath dynamics when $\R(0)$ is fixed.

Thus, the joint initial state of the overall system at $t=0$ is given by:
\BEA
\label{initial_state_2}
\rho\left(
\{ \r_k, \v_k\}\,|\,\R\,
\right)\, \rho(\R,\V),
\qquad \V=\dot\R,
\EEA
where $\rho(\R,\V)$ is arbitrary. 
}

The noise distribution is deduced from the initial state of the bath at the initial time $t = 0$. 
We assume factorized initial state of system+bath:
\begin{align}
\label{initial_state_3}
&\rho\left(
\{ \r_k, \v_k\}\right)\, \rho(\R,\V), \qquad \V=\dot\R, \\
&\rho\left(
\{ \r_k, \v_k\} 
\right) \propto 
e^{- \beta H_B+\beta\sum_{k=1}^N \Omega_k m_k\left({\r}_{k,x}\v_{k,y}-{\r}_{k,y}\v_{k,x}\right)},
\label{initial_state_4}
\end{align}
where $\rho(\R,\V)$ is the initial state of the Brownian particle, and 
where $H_B$ is the bath energy derived from \eqref{lagr_bath}. 
According to the angular momentum term in \eqref{initial_state_4}, each bath oscillator rotates around the $z$-axis located at $(x=0,y=0)$. Unlike in \eqref{general}, each oscillator has its own angular frequency $\Omega_k$; this is explained below. Both the energy $H_B$ and the angular momentum term in \eqref{initial_state_4} are conserved quantities of the bath dynamics.

The scenario of preparing (\ref{initial_state_3}, \ref{initial_state_4}) involves the bath rotating around the $z$-axis, and then the interaction with the Brownian particles is suddenly turned on. 
Eqs.~(\ref{initial_state_3}, \ref{initial_state_4}) produce the Langevin equation as with an additional force $\propto \R(0)$, which comes from (\ref{noise}, \ref{ido}). This preparation-dependent force does not influence the stationary state of the Brownian particle. For simplicity, we shall set it to zero by taking the initial conditions $\R(0)=0$.

\subsection{Stability of rotating thermal bath}
\label{stabon}

As we emphasized, each oscillator in \eqref{initial_state_4} has its own angular frequency $\Omega_k$. This is because low-frequency oscillators, while irrelevant for the bath's thermodynamics, are crucial for the long-term behavior of the Brownian particle. A constant rotation frequency $\Omega_k$ would lead to instability, 
inverting the harmonic potential. Therefore, $\Omega_k$ must decrease with $\omega_k$; see Appendix \ref{app-harmonic-gibbs}. Specifically, stability requires:
\BEA
\omega_k>\Omega_k,\quad {\rm for~all}\quad k.
\label{stability}
\EEA
The instability of low-frequency harmonic modes under rotation is a well-known phenomenon, closely related to superradiance \cite{bekenstein1998many,alicki2018interaction}. The phenomenon also relates to the electromagnetic field amplification (or generation) by a rotating body; see Ref.~\cite{zeldo} for a recent discussion and experimental verification of this effect. 


In the classical scenario, addressing these instabilities under $\Omega_k=\Omega$ will demands introducing a non-linear confining potential for the oscillators. This, combined with the inverted harmonic potential due to rotation, would create a double-well non-linear potential. Hence, the bath dynamics will no longer be exactly solvable, and the model would lose its primary advantage in exploring non-weak coupling scenarios \footnote{Another regularizing scenario is to cut off the interaction at low frequencies. This is problematic, because the interaction with low-frequency oscillators leads to relaxation at long times. Hence, cutting off these interactions will alter the relaxation behavior, and will not recover the ordinary Langevin dynamics for the non-rotating bath. }. 

Condition (\ref{stability}) is not compatible with rigid-body rotation $\Omega_k={\rm const}$ at all frequencies. However, it is compatible with the generic rotating motion of fluids \cite{acheson}. Indeed, realistic thermal baths are more similar to fluids than to rigid bodies. 
The Navier-Stokes equation predicts two cylindrical symmetric vortex solutions: the rigid-body vortex, where the angular flow velocity holds $v_\phi(r)=\Omega r$ ($r$ being the radius of the cylindrical coordinate system and $\Omega={\rm const}$), and the potential vortex, where $v_{\phi}(r)\propto r^{-1}$. Natural vortices (e.g., tornadoes) tend to interpolate between these two regimes at (resp.) small and large $r$ \cite{acheson}. Appendix \ref{app-tornado} shows that for such vortices, condition (\ref{stability}) can be satisfied.  

\subsection{The distribution of the noise}

\comment{
For simplicity we pose $\R(0)=0$ so that the additional force term in \eqref{lang} does not show up; cf.~\eqref{ido} \footnote{There is a way to avoid this $\R(0)=0$ assumption, keeping our derivations intact. To this end, we can assume that the bath with Hamiltonian ${\sum}_{k=1}^N \left[\frac{m_k}{2}\dot{\r}_k^2 +
\frac{m_{k}\omega_{k}^{2}}{2} \left(\r_{k}-\frac{c_k\R}{m_k\omega_k^2}\right)^{2}\right]$
[see \eqref{lagr_bath}] was allowed to equilibrate under a fixed value of $\R(0)$. 
Then the conditional equilibrium state of the bath will contain the orbital momentum defined via $\bar \r_k$ [see \eqref{ido}], and the assumption $\R(0)=0$ will not be needed. 
Then the initial state will read $\calP(\r_k,\v_k,\R,\V) = \calP\left(
\{ \bar \r_k, \v_k\}
\,\middle|\,
\R,\V
\right) \calP(\R,\V)$, where $\calP\left(
\{ \bar \r_k, \v_k\}
\,\middle|\,
\R,\V
\right)$ is the conditional equilibrium state of the bath, and $\calP(\R,\V)$ is an arbitrary density of the Brownian particle.
}.
}

We evaluate the noise autocorrelation function by averaging the outer product of \eqref{noise} over \eqref{initial_state_4}:
\begin{align}
    &\la \bxi(t)\tr{\bxi(t')} \ra =\,\rC_\bxi(t-t'),
    \\
\comment{    &\color{lightgray}\rC_\bxi(t) =
    \sum_k 
\left(
\begin{array}{ccc}
 \frac{T c_k^2 \cos  \left(\omega _k t\right)}{m_k \left(\omega _k^2-\Omega_k^2\right)} 
 & -\frac{T   c_k^2 \sin  \left(\omega _k t \right)}{m_k  \left(\omega _k^2-\Omega_k^2\right)} 
 \frac{\Omega_k}{\omega_k}
 & 0 
 \\
 \frac{T   c_k^2  \sin  \left(\omega _k t \right)}{m_k  \left(\omega _k^2-\Omega_k^2\right)} 
 \frac{\Omega_k}{\omega_k}
 & \frac{T c_k^2 \cos  \left(\omega _k t \right)}{m_k \left(\omega _k^2-\Omega_k^2\right)} 
 & 0 
 \\
 0 
 & 0 
 & \frac{T c_k^2 \cos  \left(\omega _k t \right)}{m_k \omega _k^2} 
 \\
\end{array}
\right)
\\
}
&\rC_\bxi(t) \!=\!\!
    \sum_k \frac{T c_k^2}{m_k}\!
\left(\!\!
\begin{array}{ccc}
 \frac{\cos  \left(\omega _k t\right)}{ \left(\omega _k^2-\Omega_k^2\right)} 
 & \!\!-\frac{   \sin  \left(\omega _k t \right)}{  \left(\omega _k^2-\Omega_k^2\right)} 
 \frac{\Omega_k}{\omega_k}
 & 0 
 \\
 \frac{ \sin  \left(\omega _k t \right)}{ \left(\omega _k^2-\Omega_k^2\right)} 
 \frac{\Omega_k}{\omega_k}
 & \frac{ \cos  \left(\omega _k t \right)}{\left(\omega _k^2-\Omega_k^2\right)} 
 & 0 
 \\
 0 
 & 0 
 & 
 \!\!\!\!\!\!\!\frac{\cos  \left(\omega _k t \right)}{\omega _k^2} \!\!
\end{array}
\right)\!.\!\!\!
\label{prm}
\end{align}
where $T=1/\beta$. Assuming a dense spectrum of oscillators with frequencies defined as $\omega_k = \delta \omega k$, where $\delta \omega \to 0$, we can approximate the summation in \eqref{prm} by integration. To this end, we define $c(\omega_k)=c_k/\sqrt{\delta\omega}$ (the interaction with each single oscillator is weak), $m(\omega_k) = m_k$, and $\Omega(\omega_k) = \Omega_k$ for positive arguments. For negative arguments $\omega<0$, we define
\begin{equation}
\label{kako}
c(\omega) \!=\! c(-\omega), \quad m(\omega) \!=\! m(-\omega), \quad \Omega(\omega)\!=\!-\Omega(-\omega),~~
\end{equation}
We also express $\sin$ and $\cos$ using complex exponentials, extending the integration range to $(-\infty, \infty)$:
\begin{equation}
\label{C-gen}
\begin{aligned}
&\rC_\bxi(t) = \frac{1}{2\pi}
\int_{-\infty}^\infty \!\!\!\!\d\omega\, \ee^{\ii \omega t}\; \frac{\pi T c(\omega)^2}{m(\omega)}
\,\times
\\
&\qquad\times
\left(
{
\begin{array}{ccc}
\frac{1}{\left(\omega^2-\Omega(\omega)^2\right)}
 & \tfrac{\ii\,\sgn(\omega)}{\left(\omega^2-\Omega(\omega)^2\right)} 
 \tfrac{\Omega(\omega)}{\omega} & 0 
 \\
\tfrac{-\ii\,\sgn(\omega)}{ \left(\omega^2-\Omega(\omega)^2\right)} 
 \tfrac{\Omega(\omega)}{\omega}
 & 
 \frac{1}{\left(\omega^2-\Omega(\omega)^2\right)}
 & 0 
 \\
 0 
 & 0 
 & \tfrac{1}{\omega^2} 
 \\
\end{array}
}
\right).
\end{aligned}
\end{equation}
where $\sgn(\omega)$ is the sign function.
The spectral density $\rS_{\bxi}(\omega)$ is the Fourier transform of the autocorrelation function $\rC_\bxi(t)$
\footnote{\label{ff1}We use for the Fourier transform: $\td{f}_\omega = \int_{-\infty}^\infty f(t) \ee^{-\ii \omega t} \, \d t$.}. Eq.~\eqref{C-gen} implies:
\begin{equation}
    \label{S}
\begin{aligned}
&\rS_{\bxi}(\omega)= \frac{\pi T c(\omega)^2}{m(\omega)}
\times\\
&\qquad\times
\left(\!\!
\begin{array}{ccc}
\frac{1}{\left(\omega^2-\Omega(\omega)^2\right)}
 & \tfrac{\ii\,\sgn(\omega)}{\left(\omega^2-\Omega(\omega)^2\right)} 
 \tfrac{\Omega(\omega)}{\omega} & 0 
 \\
\tfrac{-\ii\,\sgn(\omega)}{\left(\omega^2-\Omega(\omega)^2\right)} 
 \tfrac{\Omega(\omega)}{\omega}
 & 
 \frac{1}{\left(\omega^2-\Omega(\omega)^2\right)}
 & 0 
 \\
 0 
 & 0 
 & \tfrac{1}{\omega^2} 
 \\
\end{array}\!\!
\right).
\end{aligned}
\end{equation}
In the continuum limit, the memory kernel \eqref{kernel} reads:
\begin{align}
\label{zeta-gen}
    &\zeta(t) = \int_{-\infty}^\infty \frac{c(\omega)^2}{2m(\omega) \omega^2} \ee^{\ii \omega t} \d\omega.
\end{align}
Notice when $\Omega(\omega)\equiv0$, the autocorrelation \eqref{C-gen} simplifies to $\rC_\bxi(t) = T \zeta(t) \rI$, where ($\rI$ is the identity matrix). This implies the fluctuation-dissipation relation \cite{seifert2012stochastic,landau_stat,tuckerman,breuer2002theoryofopenquantum}. 

Choosing the coupling coefficients as \cite{magalinskii1959dynamical,breuer2002theoryofopenquantum,zwanzig1973nonlinear,caldeira1983quantum,tuckerman}
\begin{align}
    \label{c-white}
    \frac{c(\omega)^2}{m(\omega)} = \frac{2\gamma\omega^2}{\pi},
\end{align}
the memory kernel \eqref{zeta-gen} simplifies to
\begin{align}
    \zeta(t) = 2 \gamma\,\delta(t),
\label{kerner-local}
\end{align}
which corresponds to local (Ohmic) friction with magnitude $\gamma$ in \eqref{lang}.

\begin{figure}[t]
\centering
\includegraphics[width=0.95\linewidth]{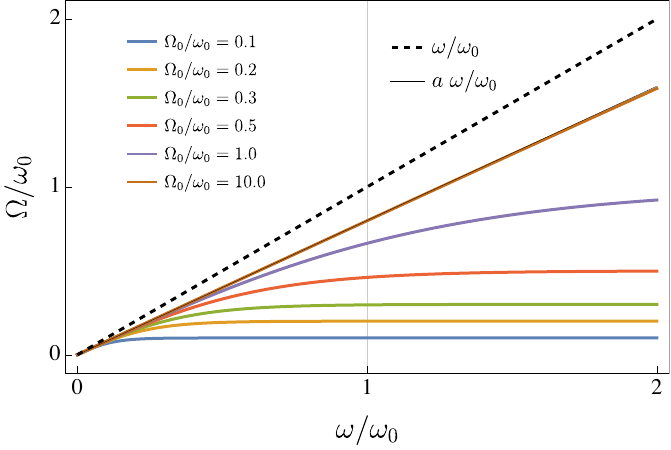}    
\caption{
Visualisation of the angular velocity spectrum of the oscillators \eqref{omega-tanh} with $a=0.8$ (solid lines). For large $\Omega_0$, this approaches the linear limit $a\,\omega$ in \eqref{linear}.
}
\label{fig-omegas}
\end{figure}

For simplicity, we select the following angular velocity spectrum of the bath that satisfies \eqref{stability}:
\begin{align}
\label{omega-tanh}
    \Omega(\omega) = \Omega_0 \tanh(a \,\omega / \Omega_0), \qquad 0 \le a < 1.
\end{align}
This function ensures that the rotation increases linearly with the oscillator frequency and saturates at $\Omega_0$.
The parameter $a$ represents the slope at low frequencies, with $0 \le a < 1$.
In Fig.~\ref{fig-omegas}, $\Omega(\omega)$ is plotted for various values of $\Omega_0$.
With this choice, the thermal bath exhibits a uniform rotating frequency for modes with macroscopically large frequencies, while for low frequencies, $\Omega(\omega) \approx a \,\omega$, as needed for stability; cf.~\eqref{stability}.
We stress that the main virtue of (\ref{omega-tanh}) is simplicity and clarity in holding (\ref{stability}). Fluid dynamics suggests a different form of $\Omega(\omega)$ (which still satisfies (\ref{stability})), but eventually leads to the same quantitative conclusions; see Appendix \ref{app-tornado}.

Using \eqref{c-white} and \eqref{omega-tanh} in \eqref{C-gen}, we obtain:
\begin{align}
&\rC_\bxi(t)_{11} = \rC_\bxi(t)_{22} 
\nonumber\\
&\qquad~=\frac{1}{2\pi}
\int_{-\infty}^\infty \!\!\!\!\d\omega\, \ee^{\ii \omega t}\; \frac{\pi T c(\omega)^2}{m(\omega)}
 \frac{1}{\omega^2-\Omega(\omega)^2} 
\nonumber\\
&\qquad~=2\gamma T\delta(t) + \frac{2\gamma T}{\pi} 
 \int_{0}^\infty \frac{\d\omega\,\cos(\omega t)\Omega(\omega)^2}{\omega^2-\Omega(\omega)^2}.
 \label{ubu}
\end{align}
We get a correction to the standard delta-correlated noise with correlation function ${2\gamma T}\delta(t)$. The correction is finite at $t=0$ and decays exponentially for $t\gg a/\Omega_0$. 
The exponential decay is confirmed numerically. It follows from the fact that if we approximate in (\ref{ubu}): $\Omega(\omega)\simeq a\omega$ for small frequencies or large times, we get $\propto \delta(t)$ for the integral in (\ref{ubu}). Likewise, we find
\begin{align}
&\rC_\bxi(t)_{21} = -\rC_\bxi(t)_{12} 
\nonumber\\
&\qquad=\frac{1}{2\pi}
\int_{-\infty}^\infty \!\!\!\!\d\omega\, \ee^{\ii \omega t}\; \frac{\pi T c(\omega)^2}{m(\omega)}
 \frac{\sgn(\omega)}{\ii \left(\omega^2-\Omega(\omega)^2\right)} 
 \frac{\Omega(\omega)}{\omega}
\nonumber \\
&\qquad= \frac{2\gamma T}{\pi} 
 \int_{0}^\infty \frac{\d\omega\,\sin(\omega t)\omega\Omega(\omega)}{\omega^2-\Omega(\omega)^2}.
\label{gogi}
\end{align}
For $t\gg a/\Omega_0$, this expression simplifies to:
\BEA
\label{dogi}
\text{Eq. } \eqref{gogi} \simeq \frac{2\gamma Ta }{\pi(1-a^2)}\, t^{-1},
\EEA
which follows from approximating $\Omega(\omega)\simeq a\omega$ within the dominant range of integration.
In this way, we find (\ref{dogi}) multiplied by integral ${\rm lim}_{\varepsilon\to 0+}\int_{0}^\infty \d\omega\,e^{-\omega\varepsilon}\sin(\omega)={\rm lim}_{\varepsilon\to 0+}{\rm Im}\int_{0}^\infty \d\omega\,e^{-\omega\varepsilon +i\omega }=1$.
Eq.~\eqref{dogi} shows that the non-diagonal part of the noise is long-range correlated for large times. 

For $t\ll a/\Omega_0$, we get a finite expression:
\BEA
\label{bogi}
\text{Eq. } \eqref{gogi} \simeq {\gamma T }\Omega_0,
\EEA
where \eqref{bogi} is found by changing the variable $\omega t\to \omega$ in \eqref{gogi}, and approximating $\Omega(\omega/t)\simeq \Omega_0$ for $t\to 0$; cf.~(\ref{omega-tanh}).

\section{Langevin equation and the stationary state for solvable noise}
\label{sec-aomega}

Consider the specific case of a linear angular velocity spectrum within the framework of (\ref{lang}, \ref{kernel}, \ref{c-white}):
\begin{align}
\Omega(\omega) = \omega a.
\label{linear}
\end{align}
Although this scenario is not physically realistic due to the non-constant bath angular velocity for large frequencies, it is useful because of its solvability. More specifically, the relaxation dynamics and stationary state of the Brownian oscillator can be expressed using elementary functions, allowing us to illustrate features that are also observed in the more realistic case \eqref{omega-tanh}. Eqs.~(\ref{linear}, \ref{S}, \ref{c-white}) imply for the noise correlation functions:
\begin{align}
    \rS_\bxi(\omega) = 
    2\gamma T
\left(
{
\begin{array}{ccc}
\frac{1}{1-a^2}  & \tfrac{\ii a \;\sgn(\omega)}{1-a^2}  & 0 
 \\
\tfrac{-\ii a \;\sgn(\omega)}{1-a^2} & \frac{1}{1-a^2}  & 0 
 \\
 0 & 0  & 1
\end{array}
}
\right),
\end{align}
which corresponds to the autocorrelation function:
\begin{align}
\label{C-1}
    \rC_\bxi(t) =2\gamma T
\left(
\begin{array}{ccc}
 \delta(t) \frac{1}{1-a^2}
 &  \frac{-a~}{1-a^2}\frac{1}{\pi}\calP\left(\frac{1}{t}\right)
 & 0 
 \\
 \frac{a}{1-a^2} \frac{1}{\pi} \calP\left(\frac{1}{t}\right)
 & \delta(t) \frac{1}{1-a^2}
 & 0 
 \\
 0 
 & 0 
 & \delta(t) 
 \\
\end{array}
\right)\!,
\end{align}
which also directly follows from \eqref{C-gen}. Here, $\calP$ denotes the Cauchy Principal Value (CPV); see Appendix \ref{app-cauchy} for more details. Thus, any occurrence of ${1}/{t}$ in an integral should be understood in the sense of CPV. To derive \eqref{C-1}, we used \eqref{dodosh} from Appendix \ref{app-cauchy}.

Next, we consider the harmonic potential for the system, given by $V(\R) = \frac{1}{2}\omega_0^2 \R^2$, and demonstrate the convergence to the stationary state distribution. In the absence of a magnetic field, we focus on the dynamics in the $x$-$y$ plane (orthogonal to the rotation axis $\o$, with $\tr{\R\pp}\equiv (X,Y)$). The Langevin equation \eqref{lang} reads:
\begin{align}\label{lang-harmonic}
    &\ddot\R\pp = -\omega_0^2 \R\pp -\gamma \dot\R\pp + \bxi\pp,
    \\
    \label{noise-corr}
    &\la \bxi\pp(t) \tr{\bxi\pp}(t')\ra = \frac{2\gamma T}{1-a^2}
\left( 
\begin{array}{ccc}
 \delta(t-t')
 & - \calP \frac{a}{\pi (t-t')}
 \\
 \calP \frac{a}{\pi (t-t')} 
 & \delta(t-t'). 
 \\
\end{array}
\right).
\end{align}
The separate dynamics of $X(t)$ and $Y(t)$ follow Langevin equations with white noise that refers to an effective temperature $T/(1-a^2)$. The noise correlation matrix is rotation-invariant. However, $X(t)$ and $Y(t)$ are not independent due to the off-diagonal terms in the noise correlation \eqref{noise-corr}.

\begin{figure}
\centering
\includegraphics[width=0.85\linewidth]{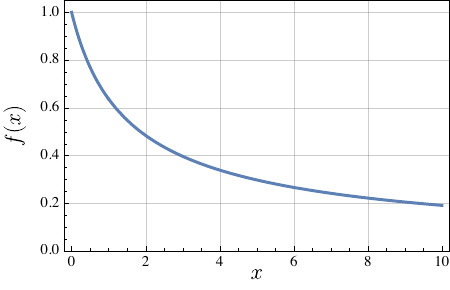}~~
\caption{Plot of function $f(x)$ defined in \eqref{f-def}.} 
\label{fig-f}
\end{figure}

For simplicity, we assume $\R\pp(0) = \dot\R\pp(0) = 0$ and apply the Laplace transform to (\ref{lang-harmonic}):
\begin{align}
    \hat\R\pp(s) =& \hat{K}(s) \hat\bxi\pp(s),
        \quad
    \hat{K}(s) = \frac{1}{\omega_0^2 + s^2 + \gamma s},
    \label{K-hat}
\end{align}
where $\hat{K}(s)$ is the Laplace transform of $K(t)$. The inverse Laplace transform of $\hat{K}(s)$ is:
\begin{gather}
\label{karp}
    K(t) = \frac{1}{\beta-\alpha}\left(\ee^{-\alpha t} - \ee^{-\beta t}\right),
\\
\label{alphabeta}
\alpha = \frac{\gamma}{2} -\frac{1}{2}\sqrt{\gamma^2-4\omega_0^2}~,
    \quad
    \beta = \frac{\gamma}{2} +\frac{1}{2}\sqrt{\gamma^2-4\omega_0^2}~.
\end{gather}
Here, $1/\alpha$ and $1/\beta$ are the relaxation times of the Brownian oscillator.
For strong coupling ($\gamma \gg \omega_0$), we have $\alpha = \omega_0^2/\gamma$ and $\beta = 1/\gamma$, indicating significantly different relaxation times (overdamped limit).
Eqs.~(\ref{lang-harmonic}, \ref{K-hat}) yield
\begin{align}\label{RKxi}
    \R\pp(t) =& \int_0^t K(t-u)\bxi\pp(u)\,\d u.
\end{align}
In Appendix \ref{app-inteval} we evaluate $\la \R\pp\dot\R\pp\tr{}\ra$
and other moments in the limit $t\to\infty$. The only non-trivial moment is
\begin{align}
    \la X V_y\ra
    =&\; 
    T\frac{a\omega_0}{  \omega_0^2 - (a\omega_0)^2}
    \; f\left(\frac{\gamma}{2\omega_0}\right)
    = -\la Y V_x\ra,
\label{xvy-line3}
\end{align}
where $f(x)$ is a smooth, monotonically decreasing function (see Figure \ref{fig-f}) defined as:
\begin{equation}
\label{f-def}
\begin{aligned}
    &f(x)=\begin{cases}
    \frac{1}{\pi\,\sqrt{x^2-1}}\log\left(
    \frac{x+\sqrt{x^2-1}}{x-\sqrt{x^2-1}}
    \right) \qquad &\text{when}~~ x>1,
    \\[2ex]
    \frac{2 \arctan\left(\sqrt{1-x^2}/x\right)}{\pi\,\sqrt{1-x^2}}\qquad &\text{when}~~ x<1,
    \end{cases}
    \\
&f(0)=\,1,\qquad f(1)=\frac{2}{\pi}.
\end{aligned}
\end{equation}

Since the dynamics of $X$ alone and $Y$ alone in \eqref{lang-harmonic} follow standard Langevin equations with white noise, we obtain:
\begin{align}
\label{upo}
    \la X^2 \ra = \la Y^2 \ra = \frac{T}{\omega_0^2 - (a\omega_0)^2},
    \\
    \la V_x^2 \ra = \la V_y^2 \ra = \frac{T \omega_0^2}{\omega_0^2-(a\omega_0)^2}.
\label{upo2}
\end{align}
All other second-order moments vanish. The Gaussian generated by (\ref{xvy-line3}, \ref{upo}, \ref{upo2}) is matched with the rotating Gibbs distribution \eqref{general} for the harmonic potential; see Appendix \ref{app-harmonic-gibbs}. This amounts to introducing effective temperature $T_\eff$ and effective angular velocity $\Omega_\eff$:
\begin{align}
    \label{xx}
    T_\eff =T \frac{\omega_0^2-\Omega_\eff^2}{\omega_0^2 - (a\omega_0)^2},\qquad
    \Omega_\eff =a\,\omega_0\; f\left(\frac{\gamma}{2\omega_0}\right).
\end{align}
Since $f\left(\frac{\gamma}{2\omega_0}\right) \leq 1$ from \eqref{f-def}, it follows that $T_\eff$ is always greater than the bath temperature $T$, while $\Omega_\eff$ is always less than $a\omega_0$. 
The latter is the angular velocity $\Omega_k$ for the bath oscillator with the frequency $\omega_k=\omega_0$.
In the weak coupling limit $\gamma \to 0$, we have $T_\eff = T$ and $\Omega_\eff = a\,\omega_0$.
For strong coupling ($\gamma \to \infty$), $\Omega_\eff \to 0$ and $T_\eff \to T/(1-a^2)$, indicating that the Brownian oscillator's rotation ceases and its effective temperature exceeds the bath temperature.

Let us show that $T_{\rm eff}$ and $\Omega_{\rm eff}$ from (\ref{xx}) hold the Clausius equality, i.e., they have thermodynamical meaning. For clarity, we write the stationary probability again:
\begin{align}
\rho_{\rm st}(\R,\V)\propto {\rm e}^{-\frac{1}{T_{\rm eff}}[ \frac{\V^2}{2}-\o_{\rm eff}\cdot \R\times \V +\frac{\omega_0^2}{2}|\R|^2)  ]}.
\label{tutu1}
\end{align}
Under a slow variation of the parameter $\omega_0$ we get for the mean energy
\begin{align}
\frac{{\rm d} U}{{\rm d}\omega_0 }=
\frac{\partial}{\partial \omega_0} 
\int\d^3R\,\d^3V \rho_{\rm st}(\R,\V)H(\R,\V),
\label{hausa}
\end{align}
where $H(\R,\V)=\frac{1}{2}(\V^2+\omega_0^2\R^2)$. For the work we get
\begin{align}
\frac{\delta W}{{\rm d}\omega_0 }=
\int\d^3R\,\d^3V \rho_{\rm st}(\R,\V)\frac{\partial H}{\partial \omega_0}. 
\end{align}
For a review of the standard notion of work, see Section~\ref{reminder}. Energy and work define heat via the first law
\begin{align}
    {\rm d} 
&    U=\delta Q+\delta W,  \\
&
\frac{\delta Q}{{\rm d}\omega_0 }=
T_{\rm eff} 
\frac{\partial }{\partial \omega_0} 
[-\int\d^3R\,\d^3V \rho_{\rm st}\ln \rho_{\rm st}] \nonumber \\
&+ \o_{\rm eff} \cdot \frac{\partial }{\partial \omega_0} \int\int\d^3R\,\d^3V 
\rho_{\rm st}\bm{L},
\end{align}
where we employed the fact that anglar momentum $\bm{L} = \R\times\V$ does not depend on $\omega_0$. Altogether, we get the Clausius equality:

\begin{align}
\label{ada8}
\delta Q= T_{\rm eff}\, {\rm d} S + \o_{\rm eff}\cdot {\rm d} \langle\bm{L}
\rangle.
\end{align}
Note however that the transition to the free-energy from (\ref{ada8}) is not straightforward, since 
$T_{\rm eff}$ and $\o_{\rm eff}$ in (\ref{ada8}) depend on $\omega_0$. 

Thus, we conclude that strong coupling with the bath prevents the establishment of the rotating Gibbs equilibrium at the same temperature as the bath. Recall that for a non-rotating bath, i.e., $\Omega(w) = a = 0$, we do get the non-rotating Gibbs distribution from the Langevin equation (\ref{lang-harmonic}). This fact is seen directly from \eqref{xx}, and it holds for anharmonic confining potentials.
These conclusions are obtained for bath angular velocities \eqref{linear}, which is less realistic than (\ref{omega-tanh}). Below, we show that similar conclusions can also be drawn for (\ref{omega-tanh}).

In the quantum domain (i.e., for the quantum Langevin equation), a similar conclusion holds: the quantum Gibbs distribution is not achieved for a quantum Brownian oscillator strongly coupled to a quantum thermal bath \cite{an}. In the quantum case, angular momentum is not involved, and the reason for not establishing the proper Gibbs distribution at a non-weak coupling is the quantum entanglement between the system and the bath.
Classically, the angular momentum is involved, and the lack of the rotating Gibbs distribution \eqref{general} is due to specific features of angular momentum transfer between the bath and the Brownian particle. Note in this context that the angular momentum is more additive than energy, since (in contrast to the concept of interaction energy), there is no concept of ``interaction angular momentum,'' at least in non-relativistic physics.

\section{General Angular Velocity Spectrum of the Bath}
\label{sec-ft}

\subsection{Stationary State}
\label{stato}

In Section \ref{sec-aomega}, we demonstrated the convergence to the stationary state distribution despite the presence of long-range correlations in the noise using the Laplace transform. In this section, we focus on analyzing the stationary state. The relaxation times for this stationary state are still governed by \eqref{alphabeta}. To evaluate the stationary state quantities, we employ the Fourier transform method, which enables us to consider more general forms of $\Omega(\omega)$ in \eqref{C-gen}. 
Consider the Fourier transform of \eqref{lang} [cf.~Footnote~\ref{ff1}]:
\begin{align}
    \label{ft-enq}
\tilde{\R}\pp{}(\omega) =&\, \rM(\omega) \tilde{\bxi}\pp(\omega),
\\
     \rM(\omega) =& \begin{pmatrix}
        \omega_0^2-\omega^2+\ii \gamma \omega \!\! & \ii b\omega\\
        -\ii b\omega & \!\!\omega_0^2-\omega^2+\ii \gamma \omega
    \end{pmatrix}^{-1}.
    \label{ft-M}
\end{align}
Let $\rS_{\bxi\pp}(\omega)$ and $\rS_{\R\pp}(\omega)$ be Fourier transforms of (resp.) autocorrelation functions $C_{\bxi\pp}(t) = \langle \bxi\pp(0)\tr{\bxi\pp}(t)\rangle$ and $\rC_{\R\pp}(t) = \la \R\pp(t) \tr{\R\pp}(0)\ra$; cf. \eqref{S}. Using the Wiener-Kinchin theorem for stationary processes (see Appendix \ref{app-wiener}):
\begin{align}
\langle \tilde{\bxi}\pp(\omega) \tr{\tilde{\bxi}\pp}(\omega')\rangle = 2\pi \delta(\omega + \omega') \rS_{\bxi\pp}(\omega),
\end{align}
we obtain the relation:
\begin{align}
    \rS_{\R\pp}(\omega) = \rM(\omega) \rS_{\bxi\pp}(\omega) \rM(-\omega)\tr{}.
    \label{SR-M}
\end{align}
Given the definition of $\rS_{\R\pp}(\omega)$, the second moments are expressed as
\begin{eqnarray}
\label{RR}
    \la \R\pp \tr{\R\pp}\ra &\equiv& \rC_{\R\pp}(0) = \frac{1}{2\pi}\int_{-\infty}^\infty\!\!\! \rS_{\R\pp}(\omega)\,\d\omega~~~\\
\label{dotRR}
    \la \!\dot\R\pp \tr{\R\pp}\ra  &=& ~~\frac{\d}{\d t}\rC_{\R\pp}(t)\Big|_{t=0} \;= \frac{1}{2\pi}\int_{-\infty}^{\infty}\!\!\!\ii\omega \rS_{\R\pp}(\omega) \,\d\omega,~~~~~
    \\
    \la \!\dot\R\pp \tr{\dot\R\pp}\ra  &=& \!-\frac{\d^2}{\d t^2}\rC_{\R\pp}(t)\Big|_{t=0}
    \!=\frac{1}{2\pi}\int_{-\infty}^{\infty}\!\!\!\omega^2 \rS_{\R\pp}(\omega) \,\d\omega.~~~~~
\end{eqnarray}
Below, we evaluate these integrals in various regimes.

\subsection{Weak Coupling Case}
\label{sec-weak-coupling}

\begin{figure*}[t]
\includegraphics[width=\linewidth,trim=0 2ex 0 0,clip]{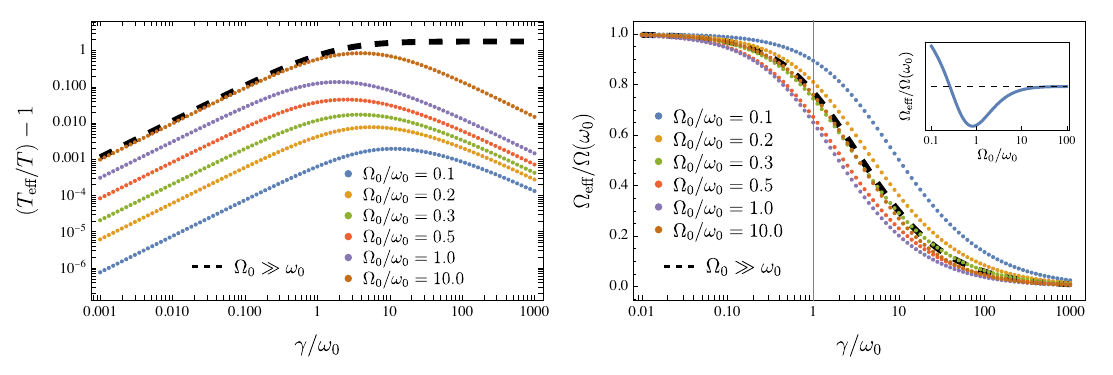}  
\caption{Given (\ref{integrals-1}--\ref{integrals-3}), we derive the effective temperature $T_{\rm eff}$ (left figure) and effective angular velocity $\Omega_{\rm eff}$ (right figure) for the Brownian oscillator using the formulas in Appendix \ref{app-harmonic-gibbs}. The dotted lines represent numerical evaluations with the nonlinear angular velocity spectrum \eqref{omega-tanh} with parameters $\Omega_0$ and $a$, while the dashed line represents the analytically solvable case of linear spectrum \eqref{linear} with $a=0.8$. We note that for $\Omega_0\gg\omega_0$, the nonlinear spectrum approaches the linear case, as also seen from Figure \ref{fig-omegas} and the plots above. Furthermore, in the weak coupling limit (low friction), the numerical estimates (dotted lines) converge to the corresponding parameters of the thermal bath (dashed line), i.e., $T^\eff\approx T$ and $\Omega^\eff \approx \Omega$, where $T$ corresponds to $1/\beta$ temperature of the thermal bath, defined in \eqref{initial_state_4}.
The ratio $\Omega^\eff/\Omega(\omega_0)$ has non-monotonic dependence on $\Omega_0$. In the inset, we fix $\gamma=\omega_0$ and plot the dependence on $\Omega_0$. The dashed line in the inset corresponds to the theoretical value for $\Omega_0 \gg \omega_0$.
It is seen that $\Omega_{\rm eff}\leq \Omega(\omega_0)$ and $T_{\rm eff}\geq T$, 
while $\Omega_{\rm eff}\to \Omega(\omega_0)$ and $T_{\rm eff}\to T$ for $\gamma\to 0$. 
}    
\label{fig-effective-params}
\end{figure*}

We consider the weak coupling limit  $\gamma\rightarrow0$ and set the magnetic field to zero; see section \ref{magno}. This makes the matrix $\rM$ in \eqref{ft-M} diagonal, and \eqref{SR-M} reduces to
\begin{align}
    \rS_{\R\pp}(\omega) = \frac{1}{|-\omega^2 + \omega_0^2 + \ii\gamma\omega|^2}\rS_{\bxi\pp}(\omega).
\end{align}
Subsequently, using (\ref{RR}, \ref{dotRR}) and (\ref{S}, \ref{c-white}) for the spectrum of the noise $\rS_\bxi(\omega)$, we express the non-vanishing moments as integrals
\begin{eqnarray}
\label{integrals-1}
    \la X^2 \ra = \la Y^2 \ra &=& \frac{1}{\pi} \int_0^\infty \frac{2  T \omega^2}{\omega^2 - \Omega(\omega)^2}\, g(\omega)\;\d\omega,~~
    \\
    \label{integrals-2}
    \la X V_y\ra = -\la Y V_x \ra &=&
    \frac{1}{\pi} \int_0^\infty \frac{2  T \omega^2\Omega(\omega)}{\omega^2 - \Omega(\omega)^2}\; g(\omega)\;\d\omega,~~
    \\
    \label{integrals-3}
    \la V_x^2 \ra = \la V_y^2 \ra &=& \frac{1}{\pi} \int_0^\infty \frac{2  T \omega^4}{\omega^2 - \Omega(\omega)^2}\, g(\omega)\;\d\omega.~~
\end{eqnarray}
where 
\begin{align}
    \label{g-nob}
    g(\omega) =  \frac{\gamma}{\gamma^2\omega^2+(\omega^2-\omega_0^2)^2}.
\end{align}

These convergent integrals reproduce the results in \eqref{xvy-line3} and \eqref{upo} when $\Omega(\omega) = a \omega$. For general $\Omega(\omega)$, the integrals can be evaluated numerically, as presented in Section \ref{sec-numeric}. 

We will further explore the weak coupling limit $\gamma \to 0$ below. The coupling coefficients become infinitely small in this limit; see \eqref{c-white}. For the integrals (\ref{integrals-1}--\ref{integrals-3}), we have 
\begin{equation}
    \frac{\gamma}{\gamma^2\omega^2+(\omega^2-\omega_0^2)^2}\rightarrow \frac{\pi}{2\omega_0^2} \delta(\omega-\omega_0)\quad {\rm for} \quad \gamma\rightarrow 0\,.
\label{nop}
\end{equation}
This is derived by noting that both sides of \eqref{nop} have the same integral $\int\d \omega...$; the latter
is calculated via contour integration and the residue theorem, noting that the four poles (for $\gamma \to 0$) are located at $\pm \omega_0 \pm i\gamma/2$. Using \eqref{nop}, \eqref{integrals-1} and \eqref{integrals-2} simplify, given that the integrand is bounded:
\begin{align}
&   \la X^2 \ra = \la Y^2 \ra = \frac{1}{\pi} \int_0^\infty \frac{2  T \omega^2}{\omega^2 - \Omega(\omega)^2}\, \frac{\pi}{2\omega_0^2} \delta(\omega-\omega_0)\;\d\omega
\nonumber    \\
&    \qquad\qquad\qquad=
    \frac{T}{\omega_0^2-\Omega(\omega_0)^2},\\
&\la X V_y\ra = -\la Y V_x \ra =\frac{T \,\Omega(\omega_0)}{\omega_0^2-\Omega(\omega_0)^2},
\\
&    \la V_x^2 \ra = \la V_y^2 \ra =\frac{T\; \omega_0^2}{\omega_0^2-\Omega(\omega_0)^2}.
\end{align}
These results agree with the rotating Gibbs distribution characterized by an average rotation frequency $\Omega(\omega_0)$ and a temperature $T$ that matches the rotating thermal bath, cf. \eqref{generalized-gibbs-moments}. 
In other words, in the weak coupling limit and without a magnetic field, the system attains the equilibrium distribution \eqref{general} with a frequency $\Omega(\omega_0)$.
Note that the stability of the system+bath is ensured by \eqref{stability}, so that $\omega_0 > \Omega(\omega_0)$. The stability holds as well for finite $\gamma$, as we saw for the exactly solvable scenario \eqref{xx}, and confirmed below numerically. 

\subsection{Numerical Investigation for Finite Friction}
\label{sec-numeric}

Using the angular velocity spectrum \eqref{omega-tanh} (visualized in Fig.~\ref{fig-omegas}), we numerically evaluate (\ref{integrals-1}--\ref{integrals-3}) for finite values of $\gamma$. Then, we compute the effective temperature and effective angular velocity of the system using these moments (cf. \eqref{pfm}). The results are presented in Fig.~\ref{fig-effective-params}. In the weak coupling limit $\gamma \to 0$, the effective parameters converge to the values imposed by the thermal bath: $\Omega_{\rm eff}\to \Omega(\omega_0)$ and $T_{\rm eff}\to T$ for $\gamma\to 0$. Fig.~\ref{fig-effective-params} shows that 
\BEA
\label{bala}
 && T_{\rm eff}\geq T,\\ 
 && \Omega_{\rm eff}\leq \Omega(\omega_0)< \Omega_0,  
 \label{ala}
\EEA
i.e., a finite coupling with the bath leads to slower rotation and a larger temperature. Note that the derivation of the Clausius inequality provided in (\ref{ada8}) applies here. 

\section{Sedimentation equilibrium and centrifugal (in)stability}
\label{sedi}

To apply our model to sedimentation [see section \ref{exoroto}], we assume that a particle of $m$ is subject to a mass-independent potential $\frac{1}{2}\chi(X^2+Y^2)$. In other words, $\chi=m\omega_0^2$ does not depend on the particle mass, since it models the external potential, which can accommodate particles of different masses. Suppose we assume the particle is immersed in a rotating thermal bath. In that case, we can apply the present model to explain how the stationary state depends on $m$, which is the main issue of sedimentation equilibrium. 

To this end, let us start from the rotating Gibbs distribution. Employing 
(\ref{generalized-gibbs-moments}) we find:
\begin{align}
\label{bodhi}
\la X^2\ra =\la Y^2\ra = \frac{T}{\chi-m\Omega_0^2}.
\end{align}
For a fixed $\chi$, $T$, and $\Omega_0$, the particle delocalization described by $\la X^2\ra =\la Y^2\ra$ indeed grows with $m$ (provided that $\chi>m\Omega_0^2$), which means that heavier particles are more likely to be found further away from the rotating axis $X=Y=0$, which is the essence of sedimentation.   

For $\chi\to m\Omega_0^2$ we witness in (\ref{bodhi}) the centrifugal instability: $\la X^2\ra =\la Y^2\ra\to \infty$. For $\chi\gtrsim m\Omega_0^2$, the relaxation time to the rotating Gibbsian state (which supports (\ref{bodhi})) is supposed to be large, while for $\chi< m\Omega_0^2$ no relaxation to the stationary state is possible, since the effective potential is no longer confining; recall (\ref{stability}).

Here are some numbers to support the applicability of (\ref{bodhi}) to real centrifuges. For a protein mass $m=10^{-19}$ g, and $\Omega_0=10^3$ Hz, we select $\chi\simeq 10^{-13}$ ${\rm g\cdot s}^{-2}$ to find $\sqrt{\la X^2\ra}\simeq 1$ cm. These are reasonable numbers for centrifuges \cite{sedimentation1,sedimentation2}. We understand that the potential of the sedimentation has to be non-linear. A clear drawback of the harmonic potential is that the probability is still maximal at the center $X=Y=0$, i.e., at the rotation axis. However, we tolerate this drawback, given the difficulty of dealing with long-range correlated noise in the presence of a nonlinear potential.

\begin{figure}[t]
    \centering
    \includegraphics[width=0.9\linewidth]{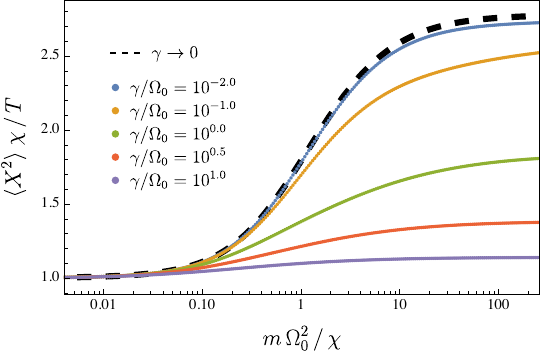}
    \caption{Sedimentation of particles of different masses coupled to a rotating thermal bath. Dependence of (scaled) mean squared displacement on the (scaled) mass of the particle; see (\ref{obukh}, \ref{omega-tanh}).}
    \label{fig-sedimentation}
\end{figure}

In our situation, we should replace (\ref{bodhi}) with 
\begin{align}
\label{bodhi2}
\la X^2\ra =\la Y^2\ra = \frac{T_{\rm eff}  }
{\chi- m \Omega_{\rm eff}^2  },
\end{align}
where $T_{\rm eff}=T_{\rm eff}|_{\omega_0=\sqrt{\chi/m}}$ and $\Omega_{\rm eff}=\Omega_{\rm eff}|_{\omega_0=\sqrt{\chi/m}}$. Now we need to study the behavior of (\ref{bodhi2}) as a function of $m$.
We do so by looking at dimensionless combinations:
\begin{align}
    \frac{\la X^2\ra \chi }{T} = \frac{T_\eff /T}{1 - \frac{\Omega_\eff^2}{\Omega_0^2}\, m\Omega^2_0/\chi}.
\label{obukh}
\end{align}
Fig.~\ref{fig-sedimentation} shows that the centrifugal instability in our situation is absent: for a large $m\Omega_0^2/\chi$, $\la X^2\ra=\la Y^2\ra$ is finite and does not diverge. We do recover the centrifugal instability discussed above only in the limit, where first $\gamma\to 0$ (and hence the rotating Gibbs is established) and only after that $m\Omega_0^2/\chi$ is taken large. 

Thus, a finite friction prevents centrifugal instability in our approach. This is related to stability conditions (\ref{stability}) imposed on the whole model. 

\section{Weak coupling under magnetic field}
\label{magno}

In the previous section, we observed that in the weak coupling limit, the system approaches the equilibrium Gibbs distribution characterized by the bath temperature $T$ and angular velocity $\Omega(\omega_0)$. In this section, we examine the impact of the magnetic field \eqref{mag} and demonstrate that, even in the weak coupling regime, the system exhibits an effective Gibbs distribution with a temperature distinct from the bath temperature.

\begin{figure}
    \centering
        \includegraphics[width=0.8\linewidth]{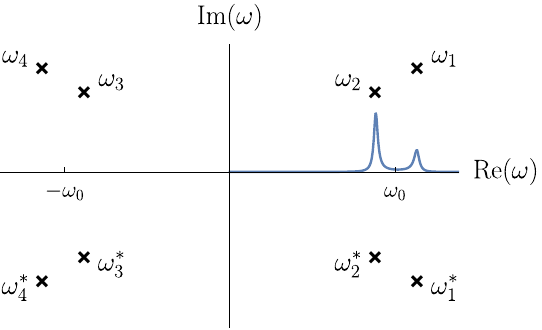}   
    \caption{This plot visualizes the poles of the function $g(\omega)$; cf.~\eqref{boro}. $0<\text{Re}(\omega_2)<\text{Re}(\omega_1)$ holds. 
    The blue curve visualizes the peaks of $g(\omega)$. 
    When $\gamma \to 0$, the roots approach the real axis, causing the corresponding peaks of $g(\omega)$ to diverge and become sharper. Consequently, only these two peaks contribute to the integrals \eqref{integrals-b-2}.
    }
    \label{fig-poles}
\end{figure}

In the presence of a magnetic field, the integrals in \eqref{RR} and \eqref{dotRR} become more complex. Detailed calculations are provided in Appendix \ref{app-magnetic}, while we highlight the primary differences between the magnetic and non-magnetic scenarios here. For instance, the integral \eqref{integrals-2} becomes
\begin{align}
    \label{integrals-b-2}
    \la X V_y\ra = -\la Y V_x \ra =&\int_0^\infty  {\phi_2(\omega)}{g(\omega)}\;\d\omega,
\end{align}
where $\phi_2$ is a smooth function, and $g(\omega)$ exhibits peaks at small $\gamma$ limit that significantly contribute to the integral:
\begin{align}
    \frac{1}{g(\omega)} =\prod_{\hat\beta=\pm b\pm \ii \gamma} 
    (\omega^2-\hat\beta\omega-\omega_0^2) 
=\prod_{i=1}^{4} 
    (\omega-\omega_i)(\omega-\omega^\ast_i),
\label{boro}
\end{align}
where $\hat\beta=\pm b\pm \ii \gamma$ takes four distinct values. The full definition of these integrals is provided in Appendix \ref{app-magnetic}. The denominator of $g(\omega)$ is an eighth-degree polynomial with eight roots; see Fig. \ref{fig-poles}.

When $\gamma$ tends to zero, these roots approach the real axis in the complex plane. Consequently, the function $g(\omega)$ has peaks at these locations. For positive $\omega$ [cf.~(\ref{integrals-1}--\ref{integrals-3}) and their analogs for the magnetic field that include only integration over $\omega\geq 0$], only the roots $\omega_1$, $\omega_2$, and their complex conjugates primarily contribute to the rapid variation of $g(\omega)$. For $\gamma \to 0$, the real parts of these roots are defined as [cf.~\eqref{mag}]:
\begin{align}
\label{loki}
\omega_\pm=\sqrt{\omega_0^2 + b^2/4}\pm b/2.
\end{align}
Compared to the form of $g$ without the magnetic field (as given in \eqref{g-nob}), the magnetic field causes the single peak to split into two peaks centered around $\omega_0$. 
Employing (\ref{generalized-gibbs-moments}) from Appendix \ref{app-harmonic-gibbs}, we show in
Appendix \ref{app-magnetic} that the stationary distribution of the Brownian oscillator subject to an external magnetic field can be mapped to the rotating Gibbs distribution (i.e. the analog of \eqref{general})
\begin{align}
&    \rho(X,Y,V_x,V_y)\propto \nonumber\\
&\exp\Big[-\frac{1}{2T_\eff} \Big (V_x^2+V_y^2+\omega_0^2 [X^2+Y^2]\Big)\nonumber\\
&+
\frac{\Omega_\eff}{T_\eff} \Big( XV_y - YV_x
        + \frac{b}{2}[X^2 + Y^2]   \Big)    \Big],
\label{thor}
\end{align}
where $XV_y - YV_x$ is the angular momentum of the Brownian particle, while $\frac{b}{2}[X^2 + Y^2]$ 
is the contribution of the magnetic field; cf.~\eqref{momentum}. The effective parameters in \eqref{thor} read 
in the weak-coupling situation $\gamma\to 0$ [cf.~\eqref{loki}]:
\begin{subequations}
    \label{eff-withb}
\begin{align}
    \label{eff-withb-1}
    T_\eff =&\, T \,\frac{\omega_+ + \omega_-}{\omega_+ + \omega_- + \Omega(\omega_+)-\Omega(\omega_-)},
    \\
    \Omega_\eff =&\, \frac{\omega_+ \Omega(\omega_-)+ \omega_- \Omega(\omega_+)}{\omega_+ + \omega_- + \Omega(\omega_+)-\Omega(\omega_-)}.
        \label{eff-withb-2}
\end{align}
\end{subequations}
Thus, the stationary state is non-Gibbsian ($T\not=T_{\rm eff}$) even in the weak coupling limit.  
The existence of $T_\eff$ and $\Omega_\eff$ is non-trivial because three moments of the Gauss density in \eqref{thor} are fit with two parameters. Since $\Omega(\omega)$ is a non-decreasing function of $\omega$ [cf.~(\ref{omega-tanh})], we conclude that $\Omega(\omega_+)\geq \Omega(\omega_-)$ from \eqref{loki}. Suppose
we are in the saturated region of the spectrum, $\Omega(\omega)\approx \Omega_0$ [cf.~(\ref{omega-tanh})], also because the magnetic field is weak. Then
$\Omega(\omega_+)\approx\Omega(\omega_-)\approx\Omega_0$, and we get from \eqref{eff-withb}: $T\approx T_\eff$ and $\Omega_0\approx\Omega_\eff$, i.e., we effectively recover the rotating Gibbs distribution under the external magnetic field.

As expected, the Bohr-van Leeuwen theorem does not hold here (see Appendix \ref{app-bvl} for discussion),
and the stationary density of the rotating particle feels the magnetic field. However, there is a difference between \eqref{eff-withb} and the no-magnetic field situation described by \eqref{xx} and Figs.~\ref{fig-effective-params}. For the latter situation, we always have $T=T_\eff$ in the weak-coupling situation $\gamma\to 0$. When the magnetic field is present and $\Omega(\omega_+)> \Omega(\omega_-)$, we get 
\BEA
T_{\rm eff}(b)<T_{\rm eff}(b=0)=T.
\label{vasil1}
\EEA
in the weak-coupling limit, as (\ref{eff-withb-1}) shows. Hence, the magnetic field leads to an effective cooling of the system. The effective temperatures under $b=0$ and $\gamma\not\to 0$ show the opposite effect, since they are larger than the bath temperature $T$; cf.~Figs.~\ref{fig-effective-params} and (\ref{bala}).

As checked numerically, we get from (\ref{eff-withb-2})
that a finite magnetic field makes the rotation of the Brownian particle slower:
\BEA
\Omega_{\rm eff}(b)<\Omega_{\rm eff}(b=0)=\Omega(\omega_0).
\label{vasil2}
\EEA
For a linear rotating spectrum $\Omega(\omega) = a \omega$, \eqref{eff-withb} simplifies:
\begin{align}
    T_\eff =\frac{T \sqrt{b^2+4\omega_0^2}}{a b+\sqrt{b^2+4 \omega_0^2}},
    \quad
    \Omega_\eff = \frac{2 a \omega_0^2}{a b+\sqrt{b^2+4 \omega_0^2}},
\end{align}
which are consistent with (\ref{vasil1}, \ref{vasil2}).

\section{Work and free-energy: rotation invariant situation}
\label{thermo}

A focus of the paper is on slow (thermodynamic) work in the stationary state, and the existence of a potential for this work, i.e., free-energy.
The general definition of work will be commented relatively briefly. 

\subsection{Free-energy without magnetic field: reminder}
\label{reminder}

The system+bath is thermally isolated and is described by the overall time-dependent Hamiltonian:
\begin{align}
\label{toto}
    H(t) = H_{S}(\lambda(t)) + H_\text{int} + H_B,
\end{align}
where the separate terms refer to (resp.) the system, interaction, and bath, and 
where $\lambda(t)$ is an externally varying parameter (external field) that directly influences the system's Hamiltonian. The work $W$ is defined as 
\begin{align}\label{W-def1}
    \frac{\delta W}{\delta t}
    = \la \pd{H}{\lambda}\dot\lambda\ra = \la \pd{H}{\lambda}\ra \dot\lambda,
    \quad \dot\lambda=\frac{{\rm d} \lambda}{{\rm d} t},
\end{align}
where $\langle...\rangle$ denotes the average over the (generally time-dependent) distribution of the system+bath; $\dot\lambda$ is taken out of the average, since it is an externally controlled variable that does not depend on the coordinates and velocities of the system+bath. For the considered thermally isolated situation, and 
using the Liouville equation of motion for the system+bath, one relates the work to the total energy change of the system+bath \cite{minima}: 
\begin{align}\label{W-def11}
    \frac{\delta W}{\delta t} = \frac{\d}{\d t} \la {H}(\,{\lambda}(t)\,)\ra.
\end{align}
Since $\pd{}{\lambda}H=\pd{}{\lambda}H_{S}$, the work in (\ref{W-def11}) can be expressed solely in terms of the system's properties:
\begin{align}
\label{W-def2}
    \frac{\delta W}{\delta t} = \la \pd{H_{S}}{\lambda}\ra \dot\lambda ,
\end{align}
where the average is now taken over the distribution of the system, which is the marginal distribution of the system+bath. Note that while the work is equal to the energy change of the total system+bath, it is not anymore equal to the energy change of the system. The difference between these two quantities is just the 
heat, which is exchanged between the system and the bath.

Eq.~\eqref{W-def1} generalizes to multiple parameters:
\begin{align}
    \label{W-def3}
\delta W = \sum_\alpha \la \pd{H_{S}}{\lambda_\alpha}\ra \d \lambda_\alpha.
\end{align}
For slow changes of $\lambda$ ($\lambda_\alpha$), the system's distribution remains close to its stationary value, which is time-dependent solely due to the dependence of $\lambda(t)$ on time. Let us now assume that this stationary distribution is given by the rotating Gibbs distribution for the system [cf.~\eqref{general} where for simplicity the rotation axis is $z$]: 
\begin{align}
\label{gibbs-drotational0}
    \rho = \frac{1}{Z(\lambda)}e^{-\beta H(\lambda)+\beta \Omega L_z}
\end{align}
where $\rho$ is the phase-space stationary density for the system (i.e., it is written in terms of canonical coordinates and momenta), $Z(\lambda)$ is the partition function. Hence, all externally changeable parameters $(\lambda_1,\lambda_2,\dots)$ 
enter into $\rho$ only via $H(\lambda)$ (and the corresponding term in $Z$). Hence, the slow work in (\ref{W-def1}, \ref{W-def3}) is written via the free-energy $F(\lambda_1,\lambda_2,\dots)$: 
\begin{align}
\label{cordoba}
&    \delta W = \d F = \sum_i \pd{F}{\lambda_\alpha} \d \lambda_\alpha,
&
     F=-T\ln Z(\lambda), \\
&    \pd{F}{\lambda_i} = \la \pd{H_{S}}{\lambda_i}\ra_{\rm eq}, & W=\Delta F,
\label{biko}
\end{align}
where $\la ... \ra_{\rm eq}$ indicates that the averaging is done via  the equilibrium density (\ref{gibbs-drotational0}), and where the final message of \eqref{biko} is that the work is given by the free-energy difference. This conclusion is important, also because it refers to the concept of thermodynamic reversibility: according to \eqref{biko}, cyclic changes of parameters not only leave the state \eqref{gibbs-drotational0}  intact (this will be the case for any stable stationary distribution), but also nullify the work. 

\subsection{Work by Time-Dependent Magnetic Field and Free Energy}
\label{freedom}

\subsubsection{General definition of work done by time-dependent and homogeneous magnetic field}

A magnetic field leads to several conceptual difficulties that should be discussed carefully. To apply (\ref{W-def11}) with $\lambda \equiv\B$, we need to introduce the Hamiltonian of a charged particle in an external (time-independent and homogeneous) magnetic field. To this end, start from \eqref{lagr_sys} and standardly define the canonic momentum $\P$, and Hamiltonian $H$ ($\tr{\R}\equiv (X,Y,Z)$):
\begin{align}
\label{ghana}
&    \P=\frac{\partial \calL_{S} }{\partial \V}=\V+\A,\quad \V=\dot\R\\
&    H=\P \cdot\V-\calL_{S}=\frac{1}{2}|\P-\A|^2+V(\R).
\label{mirabel}
\end{align}
The work is now defined by differentiating $H$ in \eqref{mirabel} over  time (entering via the time-dependent magnetic field), while $\P$ and $\R$ are held fixed [cf.~(\ref{W-def1})]:
\begin{align}
    \dot H= (\A-\P)\dot\A=-  \V \dot\A
    =-\frac{\dot\B\L}{2},\quad \L=\R\times\V,
\label{mimi}
\end{align}
where for the last equality we employed the Coulomb gauge-expression (\ref{gauge-choice}), and where $\L$ and $\frac{1}{2}\L$ are (resp.) angular momentum and magnetic moment; recall that in (\ref{lagr_sys}, \ref{mimi}) we assumed unit mass and unit charge. Eq.~\eqref{mimi} applies to non-stationary situations, where the work is defined from (\ref{W-def11}, \ref{mimi}) as 
\BEA
\delta W=-\frac{1}{2}\la \L\ra \cdot\delta \B,
\label{wowo}
\EEA
where $\la \L\ra$ is the average of the non-stationary probability density of the system. Eq.~\eqref{wowo} generalizes to an interacting system of $N$ charges $\{e_i\}_{i=1}^N$ with masses $\{m_i\}_{i=1}^N$ under homogeneous magnetic field:
\BEA
\delta W=-\sum_{i=1}^N \frac{e_i}{2} \la \R_i\times \dot\R_i\ra \cdot\delta \B,
\label{ni}
\EEA
i.e., from the viewpoint of $\delta W$ the magnetic moments add up additively \cite{landau8}, not the orbital moments $m_i \R_i\times \dot\R_i$. 

Eq.~\eqref{ni} formally coincides with the work defined by an external magnetic field in thermo-electrodynamics \cite{landau8}; see \cite{narayan} for a recent review. However, those applications refer mostly to quasi-equilibrium (slow) work, whereas \eqref{ni} applies more generally, also to non-equilibrium situations. This is relevant because a thermodynamic expression is deduced from classical Hamiltonian mechanics.   

\subsubsection{The issue of gauge-dependence}

In the context of \eqref{mimi}, we should address the following two questions. 
First, the very Coulomb gauge ${\rm div}\A=0$ is not unique for a homogeneous magnetic field $\B$: we can change 
\BEA
\A(\R)\to\A(\R)+\partial_\R\,\chi(\R), 
\label{gfree}
\EEA
where $\chi(\R)$ is a smooth function satisfying the Laplace equation $\Delta\chi=0$. Then both ${\rm rot}\A=\B$ and ${\rm div}\A=0$ are preserved, but the expression for work \eqref{mimi} will obviously change; see Ref.~\cite{giuliani2010vector} for a related discussion. This freedom of the Coulomb gauge is unphysical in the following sense. When $\B(\R)$ is inhomogeneous and decays sufficiently quickly at $|\R|\to\infty$, such behavior is also required for $\A(\R)$, and hence for $\chi(\R)$. Smooth and non-singular solutions of $\Delta\chi=0$ decaying at $|\R|\to\infty$ are zero \cite{babajan}. The homogeneous $\B$ is unphysical, e.g. because it will lead to infinite magnetic energy, since its energy density $\propto |\B|^2$ is constant. Appendix \ref{app-loop} shows that if the magnetic field $\B$ is properly regularized (so that it decays at $|\R|\to\infty$), (\ref{gauge-choice}) emerges in a region $|\R|\sim 0$, where $\B$ is approximately constant. This relates to the fact that (\ref{gauge-choice}) is linear over $\R$. Generalizations are possible, e.g., $\A(\R)=\frac{\zeta(|\R|)}{2}\B\times \R$, where $\zeta(|\R|)$ goes to zero for $|\R|\to \infty$, and $\zeta(|\R|)$ is constant for $|\R|\sim 0$. Appendix \ref{app-loop} shows that (\ref{gauge-choice}) emerges in a physical way.

The second question is more general: why should the Coulomb gauge be preferred given the gauge freedom 
\eqref{gfree} with arbitrary smooth $\chi$? This is answered in \cite{babajan}, where it is shown in the context of relativistic electrodynamics (and with the proper modeling of sources of work) that the Lorenz gauge is suitable for defining the work done by a time-dependent electromagnetic field on charged particles. For magnetic phenomena, the electric potential is neglected, and the Lorenz gauge reduces to the Coulomb gauge; see also \cite{karakhan} in this context.

\subsubsection{Free energy with magnetic field}

For slow changes of the magnetic field $\B=\ez b$, the averaging in \eqref{ni} refers to the rotating Gibbs distribution [for a single particle for simplicity; cf.~\eqref{momentum}]:
\begin{align}
\label{gge}
    \rho(\V,\R)\propto e^{-\beta[\frac{\V^2}{2}+V(\R,\lambda_\alpha)- \Omega[ (\R\times \V)_z  +\frac{b}{2}(X^2+Y^2)]]}.
\end{align}
We change to canonic variables $\R$ and $\P$ via \eqref{ghana}; note that the corresponding Jacobian is $1$:
\begin{align}
\label{bge}
    \rho(\P,\R; b,\lambda_\alpha)=\frac{1}{Z(b,\lambda)}e^{-\beta[\frac{(\P-\A)^2}{2}+V(\R,\lambda_\alpha)- \Omega (\R\times \P)_z]},
\end{align}
where the rotation axis is $\ez$ [see \eqref{mag}], and where $\A$ is given by (\ref{gauge-choice}). Eq.~\eqref{bge} 
has the form of the rotating Gibbs distribution [see \eqref{general} or \eqref{gibbs-drotational0}] 
but with the canonical orbital momentum $\R\times\P$.

For slow changes of the magnetic field, the averaging in \eqref{ni} refers to the quasi-equilibrium \eqref{bge}, i.e., to $\rho(\P,\R; b(t),\lambda_\alpha(t))$. Eq.~\eqref{ni} now refers to one component of work in \eqref{cordoba}. The same derivation as for (\ref{cordoba}, \ref{biko}) applies to \eqref{bge}, and we get that $-T\ln Z(b,\lambda)$ from \eqref{bge} is the free energy for slow changes of parameters including the magnetic field. 

We emphasize that for defining the free energy, it is necessary to have \eqref{bge}, which demands that the rotation axis and $\B$ are aligned, as already assumed in \eqref{bge}. Otherwise, the orbital momentum is not conserved, and slow changes of $\B$ do not lead to quasi-equilibrium \eqref{bge} \cite{matevosyan2024weak}; cf. our discussion before (\ref{gibbs-drotational0}). If \eqref{ni} is to be applied in the quasi-equilibrium situation, only the magnitude of the magnetic field should change (slowly) and not its direction. Hence, the free energy is defined only with respect to certain changes of $\B$. Eq.~\eqref{ni} still applies to non-equilibrium situations. 

\subsubsection{The mantra "magnetic field does no work"}

As we discussed above, work is done by a time-dependent magnetic field. However, everybody knows and frequently repeats that the magnetic field does not do (mechanical) work on charges, since the (magnetic) Lorentz force acting on any given charge is perpendicular to the velocity of that charge \cite{landau2}. It is therefore understandable that people wondered how these two statements are consistent. They are consistent, and there is an interesting quest to find microscopic mechanisms behind this consistency; see Refs.~\cite{mosca,coombes,deissler,kirk,karakhan,schmidt,babajan,ricardo} for discussions on such mechanisms.   

\subsection{Free energy of the rotating stationary state}
\label{result}

We return to the system (non-weakly) coupled to the rotating thermal bath; see sections \ref{sec-aomega}, \ref{stato}, and \ref{magno}. Now the stationary state is characterized by an effective temperature and rotation velocity; see \eqref{xx}, \eqref{eff-withb-1}, and Fig.~\ref{fig-effective-params}. These effective parameters depend on the external parameters of the system. Therefore, the standard proof for the existence of free energy outlined above cannot be directly applied. Nonetheless, Appendix \ref{app-proof} shows that the free energy does exist under the following conditions:

-- The system coupled to the rotating thermal bath model \eqref{lagr_bath} is a Brownian oscillator, i.e., the Hamiltonian $H(\R, \P;\, \lambda_\alpha)$ is a quadratic function of the canonic variables $\R$ and $\P$ for all values of the external parameters $\{\lambda_\alpha\}$ (including the magnetic field). 

-- The Hamiltonian is invariant under rotations along a fixed axis (we choose it to be $\bm{e}_z$), which is aligned with the magnetic field $\B$; see \eqref{mag}. This leads to 7 parameters $\lambda_\alpha$ that can be varied externally. The system with Lagrangian \eqref{lagr_sys} with harmonic potential $V(\R) = \frac{\omega_0^2}{2}\R^2$ is a special case of our theorem, where there are two external parameters: $\omega_0$ of the external potential and the strength of the magnetic field. We continue to maintain mass and charge as unities, since they are not easy to control externally for real Brownian particles. 

Two main differences between the present result and the free energies discussed in sections \ref{reminder} and \ref{freedom} are as follows. First, sections \ref{reminder} and \ref{freedom} addressed a general rotation invariant potential $V(\R; \lambda_\alpha)$, but were restricted to the rotating Gibbs distribution for the system. Here we consider only harmonic systems, but relax the restriction of the rotating Gibbs. Second, we only show the existence of the free energy, without providing its explicit form. However, this suffices to assert that any sufficiently slow, cyclic perturbation of the system in the parameter space $\{\lambda_\alpha\}$ results in zero net work. Consequently, work cannot be extracted from this system, despite it being only in effective equilibrium.

The main similarity of the two results is that we assume a rotation invariant system Hamiltonian. Looking for the genesis of the existence of free energy and its possible extensions, we can attempt to trace it back to the possibility that the non-Gibbsian stationary state for the Brownian oscillator (with the rotation invariant Hamiltonian) is a partial trace of the rotating Gibbs distribution \eqref{general} for the system+bath. Once the free energy exists for the rotating Gibbs, it also exists for its subsystem due to the fundamental feature \eqref{W-def2} of work. This argument does not imply any weak-coupling limit, since \eqref{W-def2} does not imply it. While the argument needs further exploration, we now move to the next section, where we relax the assumption of rotation symmetry for the system, while maintaining it for the bath, and demonstrate the emergence of a non-zero cyclic work.

\section{Work Extraction with Non-Rotation-Symmetric Potentials}
\label{sec-aniso}

In the preceding sections, we assumed that the overall system (system+bath) exhibits rotation symmetry. When the system is coupled to the thermal bath, it relaxes to an effective rotating Gibbs distribution. However, an equilibrium rotating Gibbs distribution is achieved only under specific conditions (Section \ref{sec-weak-coupling}). Additionally, we demonstrated in the previous section that, despite the effective equilibrium, the system has a free energy, which implies that no work can be extracted through a cyclic slow process.

In this section, we consider the same rotating thermal bath described by \eqref{lagr_bath}, now coupled to a Brownian particle in an non-rotation-symmetric potential:
\begin{align}
    \label{potdef}
    & V(\R;\, \varphi) = V_a\big(\sfR(-\varphi)\R\big)
    \\
    & \sfR\left(\varphi\right) = \begin{pmatrix}
        \cos\varphi & -\sin\varphi & 0\\
        \sin\varphi & \cos\varphi & 0\\
        0 & 0 & 1
    \end{pmatrix},
\end{align}
where the external parameter $\varphi$ refers to the rotation of the non-rotation-symmetric potential $V_a$ around the $z$-axis by angle $\varphi$.
Suppose $\varphi(t)$ changes over time. 
We calculate the work related to this, which is defined as:
\begin{align}
    \frac{\delta W}{\delta t} =& \la \pd{H}{\varphi} \dot\varphi\ra_\R = \la \pd{V}{\varphi} \dot\varphi \ra_\R 
\end{align}
where $\la~\ra_\R$ denotes the average over the non-stationary distribution of the system. In order to calculate the average, we write the Langevin equation (\ref{lang}, \ref{C-gen}, \ref{kerner-local}):
\begin{align}
    \label{langR}
    \ddot\R = -\partial_\R V\big(\R;\,\varphi(t)\big) - \gamma \dot\R + \bxi(t),
    \\
    \la \bxi(t)\tr{\bxi(t')}\ra = \rC_\bxi(t-t'),
\label{babo}
\end{align}
The noise spectrum is a feature of the bath and is unaffected by the asymmetry of the potential. Thus, $\rC_\bxi(t)$ is given by \eqref{C-gen}.

First, we will show that the work can be expressed in terms of the average torque exerted on the particle by the external potential:
\begin{align}
\label{palo}
    \pd{V}{\varphi} = M_z\qquad M_z = - \left[\R \times \partial_\R V(\R;\,\varphi)\right]_z
\end{align}
Then, we will show that this average does not depend on $\varphi$ when $\varphi(t)$ is a slow process. This is because the quasi-stationary distribution of the particle in the rotating frame of reference is time-independent.

Denoting $\R' = \sfR(-\varphi)\R$, we get
\begin{align}
    \partial_\R V(\R;\,\varphi) = \sfR(\varphi)\partial_{\R'} V_a\big(\R'\big)
    \\
    \partial_\varphi V(\R;\,\varphi) = \tr{\R}\partial_{\varphi}\sfR(\varphi)\partial_{\R'} V_a\big(\R'\big).
    \label{como}
\end{align}
Using $\partial_{\varphi}\sfR(\varphi) = \mathsf{P}\!\pp\sfR(\pi/2)\sfR(\varphi)$, where $\mathsf{P}\!\pp$ is the projection to the $x$-$y$ plane, we obtain from (\ref{palo}--\ref{como}):
\begin{align}
    \partial_\varphi V(\R;\,\varphi)     = \tr{\R} \,\mathsf{P}\!\pp\sfR(\pi/2)\, \partial_\R V(\R;\,\varphi) 
    \\
    =\tr{\R}\begin{pmatrix}
        0 & -1 & 0 \\
        1 & 0 & 0 \\
        0 & 0 & 0
    \end{pmatrix}\partial_\R V(\R;\,\varphi) = M_z.
    \label{harur}
    \end{align}
If $\varphi(t)$ changes slowly, the system remains in a quasi-stationary state, and the work is given by:
\begin{align}
\frac{\delta W}{\delta t} \approx \dot\varphi\la \partial_{\varphi}V \ra_{\R,\rm ss}  = \dot\varphi\la M_z \ra_{\R,\rm ss} ,
\end{align}
where the average $\la ~ \ra_{\R,\rm ss}$ is taken over the stationary distribution of \eqref{langR} with a constant $\varphi$.
Both $M_z$ and the steady-state distribution depend on $\varphi$, but this dependence will cancel out in $\la M_z \ra_{\R,\rm ss}$. To show this, recall $\R' = \sfR(-\varphi)\R$ and transform \eqref{langR} to $\R'$-coordinates:
\begin{equation}
    \label{langR1}
    \ddot{\R}' \!= -\partial_{\R'} V_a(\R')\!-\!\gamma \dot{\R}'\!+\! \bxi',
    \quad~
    \bxi'(t) = \sfR(-\varphi)\bxi(t),
\end{equation}
where $\partial_{\R} = \sfR(-\varphi)\tr{}\partial_{\R'} = \sfR(\varphi)\partial_{\R'}$. Recall that the autocorrelation matrix $\rC_\bxi(t)$ in (\ref{C-gen}, \ref{babo}) is invariant under rotations around the $z$ axis. Therefore,
\begin{equation}
\begin{aligned}
    \la \bxi'(t)\tr{\bxi'(t')}\ra 
    &= \sfR(-\varphi) \la \bxi(t)\tr{\bxi(t')}\ra 
    \sfR(\varphi)
    \\
    &= \sfR(-\varphi) \rC_\bxi(t-t') 
    \sfR(\varphi)
    =\rC_\bxi(t-t'). 
\end{aligned} 
\end{equation}
Thus, the spectrum of $\bxi'(t)$ and $\bxi(t)$ is the same in all the rotating frames.
This implies that Langevin equation \eqref{langR1} and the corresponding stationary distribution of $\R'$ do not depend on $\varphi$, in contrast to the original Langevin equation \eqref{langR}.
The torque $M_z$ in $\R'$-coordinates reads:
\begin{align}
    M_z = -\left[\R' \times \partial_{\R'} V_a(\R')\right]_z
\end{align}
where we applied $\R' = \sfR(-\varphi)\R$ to \eqref{harur}.
Therefore, $\la M_z \ra_{\R,\rm ss} = \la M_z \ra_{\R',\rm ss}$ and does not depend on $\varphi$.
The total work associated with one complete cycle of the potential, corresponding to a $2\pi$ change in $\varphi$, is given by:
\begin{align}
    \int_0^{2\pi} \frac{\delta W}{\delta \varphi} \d \varphi = 2\pi \la M_z \ra=-W_c,
\end{align}
where we introduced the minus work $W_c$, because $W_c>0$ means work-extraction, in which we are interested. 

We calculate $W_c$ for a harmonic non-rotation-symmetric potential:
\begin{align}
\label{V-aniso}
&    V_a(\R) = \frac{1}{2}\omega_x^2 X^2 + \frac{1}{2}\omega_y^2 Y^2~ + \frac{1}{2}\omega_0^2 Z^2,
    \\
&    \omega_x = \omega_0(1+\epsilon), \qquad \omega_y = \omega_0(1-\epsilon).
\end{align}
where $V_a$ has a form of an ellipse with axes matching the coordinate axes, and the external parameter $\varphi$ in \eqref{potdef} rotates the ellipse by angle $\varphi$.

The torque $M_z$ in the rotating frame evaluates to:
\begin{align}
\label{torque-1}
    \la M_z \ra = 4 \omega_0^2 \epsilon \la X'Y'\ra .
\end{align}
We evaluate $\la X' Y'\ra$ using \eqref{RR} (generalizing \eqref{ft-M} for the non-rotation-symmetric potential \eqref{V-aniso}):
\begin{align}
    &\la X' Y'\ra = \frac{1}{2\pi} \int_{-\infty}^\infty \frac{2\ii \, T \omega \Omega(\omega) \sgn(\omega)}{\omega^2 - \Omega(\omega)^2}\,\times
\nonumber    \\
    &\qquad\times\frac{\gamma}{
    (\omega_x^2-\omega^2+\ii\gamma\omega)
    (\omega_y^2-\omega^2-\ii\gamma\omega)}\;\d\omega.
\label{XY-int}
\end{align}
This integral vanishes when $\epsilon = 0$, as the integrand becomes an odd function, as $\Omega(\omega)$ is an odd function; see \eqref{kako}. 
In Appendix \ref{app-aniso}, we evaluate the integral \eqref{XY-int} in the limit $\gamma \ll \omega_0$ and $\epsilon \ll 1$, i.e., weak coupling and small asymmetry (see \eqref{xy-good}):
\begin{align}
\label{Wc-def}
    W_c =
    \frac{\epsilon^2\,\frac{\gamma}{2\omega_0}}{\epsilon^2+\left(\frac{\gamma}{2\omega_0}\right)^2} \frac{8\pi T\omega_0 \Omega(\omega_0)}{\omega_0^2-\Omega(\omega_0)^2}.
\end{align}
Note that the friction $\gamma$ must be finitely small to achieve non-zero work. For a fixed small friction $\gamma$, at weak asymmetries, the cyclic work increases quadratically with the degree of asymmetry $\epsilon$. However, it saturates at the value 
\begin{align}
    \label{Wc-max}
    W_{c,\max} =  \frac{4\pi \gamma T \Omega(\omega_0)}{\omega_0^2-\Omega(\omega_0)^2}.
\end{align}
This behavior is illustrated in Figure \ref{fig-Wc} (solid line and dashed line). 
Thus, work extraction from a single rotating thermal bath is possible via a slowly rotation asymmetric potential.

\subsection{Phenomenological Langevin Equation for a Brownian Particle Interacting with a rotating Bath}
\label{pheno}

Several phenomenological approaches have been developed to study Brownian motion in a rotating bath \cite{rotating_brownian_prl, rotating_brownian_jsp, rotating_brownian_physicaa}; see also \cite{matevosyan2024weak} for the most recent approach.
Here, a phenomenological Langevin equation was proposed that modifies only the friction term of the usual Langevin equation (${\dot\R }=\V$, unit mass):
\begin{align}
    \label{lang-pheno}
   {\dot \V}= -\partial_\R V(\R) - \gamma(\V - \o \times \R) + \sqrt{2 \gamma T} {\bxi}(t).
\end{align}
where $\R$ and $\V$ are the position and velocity of the particle, $V(\R)$ is the potential, $\gamma$ is the friction coefficient, $T$ is the temperature, and $\bxi(t)$ is the (most usual) white noise with uncorrelated components. As in \eqref{general}, the thermal bath is characterized by two parameters: the temperature $T$ and the average rotation velocity $\o$. Eq.~\eqref{lang-pheno} differs from the standard Langevin equation only in the friction term, which depends on the relative velocity between the particle and the rotating thermal bath, $\V - \o \times \R$. Ref.~\cite{matevosyan2024weak} provides phenomenological arguments for deriving \eqref{lang-pheno}, and fits this equation to numerical experiments. For rotation symmetric potentials, $V(\R) = V(|\R|)$, \eqref{lang-pheno} reproduces the rotating Gibbs distribution \eqref{general} in the long-time limit. 

The main difference between \eqref{lang-pheno} and \eqref{lang} (assuming \eqref{lang} is in the local friction limit for the sake of comparison, i.e. \eqref{kerner-local} holds) is that \eqref{lang} features long-range correlated noise with standard friction, whereas \eqref{lang-pheno} employs white noise with a modified friction term. The white-noise limit in \eqref{lang-pheno} allows for a convenient analysis using the Fokker-Planck equation for any potential $V(\R)$, unlike \eqref{lang}, where the long-range correlated noise complicates such descriptions for non-linear potentials. However, \eqref{lang} has clearer microscopic foundations.

Regardless of the initial conditions, the system described by \eqref{lang-pheno} reaches (for a confining potential) a stationary state, which for non-rotation-symmetric potential differs from the equilibrium rotating Gibbs distribution \eqref{general}. For a harmonic confining potential \eqref{V-aniso}, the exact stationary distribution of \eqref{lang-pheno} is worked out in \cite{matevosyan2024weak}. Specifically, we are interested in the cyclic work:
\begin{align}
    W_c =& -2\pi \la M_z \ra 
    = 
    \frac{\epsilon^2\,\frac{\gamma}{2\omega_0}}{\epsilon^2+
    \big(1-\frac{\Omega^2}{\omega_0^2}\big)
    \big(\frac{\gamma}{2\omega_0}\big)^2} 
    \frac{8\pi T \Omega}{\omega_0} ,
    \label{Wc-pheno}
\end{align}
where the last expression is derived under the limits $\gamma \ll \omega_0$ and $\epsilon \ll 1$. The full expression is detailed in Appendix \ref{app-pheno} and visualized in Figure \ref{fig-Wc}.

We compare the two expressions \eqref{Wc-def} and \eqref{Wc-pheno} by setting $\Omega = \Omega(\omega_0)$
(notice the phenomenological thermal bath does not have a rotation spectrum $\Omega(\omega)$). It is evident that the work extracted from our thermal bath model is larger than that from the phenomenological bath, as illustrated by the dashed lines in Figure \ref{fig-Wc}.

The solid lines representing the exact values of $W_c$ intersect at larger $\epsilon$ values (note that the solid blue line is evaluated numerically). It appears that at sufficiently large asymmetries, more work can be extracted from the phenomenological thermal bath. However, the assumption $\Omega = \Omega(\omega_0)$ in \eqref{Wc-pheno} is valid only in the weak coupling limit; see sections \ref{sec-aomega} and \ref{sec-weak-coupling}. Strictly speaking, in the non-rotation-symmetric case, the weak coupling limit is defined as $\gamma \ll \min(\omega_x, \omega_y)$, which gets violated for $\epsilon \approx 1$. 

At small asymmetries, the two expressions for cyclic work align. This alignment arises because, in the weak coupling limit and in the absence of magnetic fields, the system coupled to either thermal bath relaxes to the rotating Gibbs distribution, despite the fundamental differences in the underlying Langevin equations.

\begin{figure}
    \centering
    \includegraphics[width=0.98\linewidth]{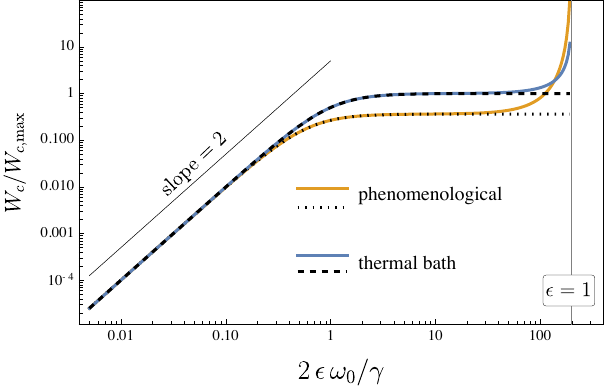}
    
    \caption{Cyclic work $W_c$ extracted from the system as a function of the asymmetry $\epsilon$. Two types of thermal baths are considered: the microscopic model of the rotating thermal bath (solid blue and dashed lines, Section \ref{sec-aniso}) and the phenomenological rotating thermal bath (solid orange and dotted lines, Section \ref{pheno}). $W_{c,\max}$ denotes the saturation value of the work for the rotating thermal bath model; cf.~\eqref{Wc-max}. The solid blue line represents the numerical evaluation of the integral \eqref{XY-int}, while the solid orange line is derived analytically in Appendix \ref{app-pheno}; see \eqref{XY-pheno-full}. The dashed and dotted lines correspond to the simplified expressions \eqref{Wc-def} and \eqref{Wc-pheno}, respectively, valid in the weak coupling ($\gamma \ll \omega_0$) and small asymmetry ($\epsilon \ll 1$) limits, explaining their deviation from the solid lines near $\epsilon = 1$. At small $\epsilon$, the cyclic work scales as $W_c \propto \epsilon^2$, but it saturates for larger $\epsilon$. For sufficiently small $\epsilon$, more work can be extracted from our rotating thermal bath compared to the phenomenological thermal bath under the same cyclic slow process. The intersection of the solid lines is discussed in the main text. Parameters used: $\gamma = 0.01$, $\omega_0 = 1$, and $\Omega(\omega) = \tanh(0.8\omega)$ (the same as the purple line in Figure \ref{fig-omegas}).}
    \label{fig-Wc}
\end{figure}

\subsection{Experimental proposal: optically trapped particles in a rotating bath} 

In order to test experimentally the rotating Brownian motion in practice, a colloidal bead 
(radius $1$--$5\,\mu$m) in water can be rotated steadily about the $z$-axis. The rotation can be realized 
by rotating the stage or using a microfluidic swirl. The bead is confined by an 
optical tweezer \cite{jones2015optical}
that realizes a near-harmonic potential, calibrated to obtain $\omega_0$ and a weak ellipticity $\epsilon$; cf.~\eqref{V-aniso}. For trapped molecules, rotation can also be implemented via electromagnetic fields (optical centrifuge) \cite{centrifuge_optical}.

A slow cyclic protocol (Section~\ref{sec-aniso}) is implemented by adiabatically rotating the trap’s weak ellipticity $\varphi(t):0\to2\pi$ (and/or weakly modulating the principal stiffnesses) with a period much longer than the relaxation times in \eqref{alphabeta}. From tracked trajectories $(X(t),Y(t))$, the cycle work is estimated via (\ref{palo}, \ref{torque-1}). The measured work $W_c(\epsilon,\gamma,\Omega(\omega_0))$ can then be compared to \eqref{Wc-def} and its saturation bound \eqref{Wc-max} testing the predicted $\epsilon^2$ scaling at small asymmetry and the role of finite friction.

\section{Summary}

Our primary goals in this paper were to analyze the microscopic, solvable, oscillator Caldeira-Leggett model for Brownian motion subject to a rotating thermal bath, and to study (mostly slow) work-exchange processes taking place in the rotating stationary state. For rotating thermal baths, the angular momentum is a relevant variable along with energy; cf.~(\ref{general}). Both are conserved, and both are additive \footnote{One argument against giving the conservation of angular momentum equal status with energy conservation is that statistical systems normally have a boundary, which generically does not have rotation symmetry; see \cite{dubrovskyi,caravelli,matevosyan2024weak} for a discussion of this issue. Here, we apply \eqref{general} to the local rotating surroundings of the Brownian particle, i.e., the global boundary for the bath is not relevant. }. 
Here are our results, along with their perspectives and limitations.

-- The advantage of the Caldeira-Leggett model is that it is solvable and allows us to deduce exactly the Langevin equation for the Brownian particle. However, the bath composed of harmonic oscillators is unstable for homogeneous rotation. This instability is due to low-frequency oscillators, which are important for the long-time behavior of the Brownian particle. 

Our recipe for preventing this instability is to assume that the angular rotation velocity varies with the frequency of the oscillator; see \eqref{stability}. An alternative (and perhaps more physical) way of resolving this problem is to assume that bath oscillators are non-linear, at least for sufficiently low frequencies. This will likely change the bath conceptually and lead to new effects (e.g., those related to superradiance \cite{bekenstein1998many,alicki2018interaction} and the Zeldovich effect \cite{zeldo}), but the model will lose solvability and will be much more difficult to analyze.  

We emphasize that our regularization scenario -- lower frequency oscillators rotate slower [cf.~(\ref{stability})] -- is physical and is generically realized in fluid dynamics vortices; see Appendix \ref{app-tornado} for details.

-- The main difference of the Langevin equation driven by the rotating bath is that the noise becomes long-range ($\sim t^{-1}$) correlated in the off-diagonal sector due to the rotation; see (\ref{gogi}--\ref{bogi}). This feature of the rotating Langevin equation will stay in more general bath models. Its conceptual importance stems from the fact that ordinary Fokker-Planck methods do not apply to the long-range correlated noise. Most likely, this should mean the rotation will lead to new effects in e.g. the particle escape from a non-linear, metastable potential. 

-- In contrast to the ordinary Langevin equation, the stationary state of the rotating Langevin equation is not the rotating Gibbs distribution \eqref{general}. This point is important to emphasize, since phenomenological approaches to the Brownian motion due to a rotating bath were proposed with an apparently desirable feature that the stationary state is the rotating Gibbs distribution; see section \ref{pheno}. The latter is recovered only for a vanishingly weak coupling with the bath.

For a finite coupling with the bath (i.e., a finite friction) the stationary state of the Brownian oscillator is characterized by an effective temperature $T_{\rm eff}$ and an effective angular velocity $\Omega_{\rm eff}$. Both these quantities have a clear thermodynamic meaning [see (\ref{tutu1}--\ref{ada8})], though they depend on all the involved parameters. However, they hold interesting inequalities; cf.~(\ref{xx}, \ref{bala}, \ref{ala}).

-- The non-Gibbsianity leads to an important prediction for the sedimentation equilibrium, since a finite friction (inherent in the non-Gibbsianity) prevents centrifugal instability; see section \ref{sedi}. 

\comment{
For a free Brownian particle which is subject to a rotating bath, the model holding \eqref{stability} predicts that for long times $t$ the rotation for the free Brownian particle scales as $\sim \ln t$, in contrast to the coordinate, which scales as $\sqrt{t}$; see Appendix \ref{freefreeefree}. Thus, for the free particle (in contrast to the bound particle we studied here) the rotation is irrelevant at large times. 
}

-- If the external potential acting on the Brownian particle is rotation symmetric, the above non-Gibbsian state inherits an important Gibbsian feature. There exists a free energy, i.e., work-exchange via cyclic, slow processes is impossible. In other words, the concept of reversibility is defined despite the rotation, long-range correlation and non-Gibbsian stationary state; see section \ref{thermo}. If the potential is not rotation symmetric, there are cyclic, slow processes that extract work. A generic class of such processes was shown in section \ref{sec-aniso}. We complement the detailed theory of this effect with an experimental proposal on trapped particles.

\comment{
It should be interesting to understand to what extent these or similar processes are involved in the work-extraction from rotating parts of ATP; see section \ref{exoroto}.}

-- A rotating charged Brownian particle is an interesting example of a classical system that feels the external magnetic field in its stationary state; see section \ref{magno}. All such systems have to be out of the standard Gibbsian equilibrium in one way or another; see \cite{matevosyan2021nonequilibrium,lasting} for recent discussions. 
A surprising point is that under a magnetic field, one needs more stringent conditions for recovering the rotating Gibbs distribution in the weak coupling limit (than with zero magnetic field); cf.~Section \ref{magno}. 

-- We also clarified the concept of work done for a time-dependent magnetic field. This was an unclear subject, but our results presented in section \ref{thermo} show that it is nearly as well defined as the ordinary work. 

\acknowledgements

We thank Varazdat Stepanyan for his interest in this work. 
The work was supported by HESC of Armenia, Grant Nos. 21AG-1C038 and 22AA-1C028.

\bibliography{refs}

\appendix

\section{Reminder on the Bohr--van Leeuwen theorem}
\label{app-bvl}

\noindent\textbf{Statement.} For a classical system in canonical equilibrium in a static homogeneous magnetic field $\B$ the equilibrium magnetization $\bm M$ vanishes: $\bm M= -\partial F/\partial \B=0$, where $F$ is the free energy \cite{vanvleck-quantum,landau_stat}.

\noindent\textbf{Proof.} The partition function reads
\begin{multline}\label{partition-intro}
Z \propto \int \prod_{i} \d^3 \R_i \, \d^3 \P_i\; \times
\\
\times\exp\Big\{-\beta\Big[\tfrac{1}{2m}|\P_i-q\A(\R_i)|^2+V(\{\R\})\Big]\Big\}.
\end{multline}
where $\R_i$ and $\P_i$ are the canonical coordinates and momenta of the $i$-th particle with charge $q$, $\A$ is the vector potential of the magnetic field $\B$ (cf. \eqref{mag}).
Perform the change of variables $\P_i\to \P_i+q\A(\R_i)$. The Jacobian is one and the integration domain is unchanged, hence the dependence on $\B$ is eliminated from the partition function. Therefore $F=-T\ln Z$ is independent of magnetic field, and $\bm M=-\partial F/\partial \B=0$ \cite{landau_stat,vanvleck-quantum}.

\noindent\textbf{The rotating case.} The above derivation relies on the canonical Gibbs distribution without additional conserved quantities. In the rotating setting, the stationary ensemble carries the conserved angular momentum in the exponent; cf.~\eqref{general}, \eqref{bge}, and see \eqref{momentum} for a non-zero magnetic field $\B$. Moreover, away from the strict weak-coupling limit, the stationary state is not Gibbsian, as seen from sections \ref{sec-aomega} and \ref{sec-ft}. The above momentum-shift argument then fails and the stationary distribution becomes $\B$-dependent; see Sec.~\ref{magno} and Eq.~\eqref{thor}.

\section{Rotating Gibbs Distribution for Harmonic Potential}
\label{app-harmonic-gibbs}

In this section, we evaluate the rotating Gibbs distribution \eqref{general} for a single particle with unit mass in a harmonic potential $V(\R) = \frac{1}{2}\omega_0^2 |\R|^2$. The angular velocity is in the $z$ direction, and $\o = \ez \Omega$. We shall additionally assume that the particle is charged and is subject to a magnetic field. Then \eqref{momentum} shows that the correct generalization of the Gibbs distribution is
\begin{align}
    &\rho(X,Y,V_x,V_y)\propto \exp\left[
    -\frac{\omega_0^2}{2T}\left(X^2+Y^2\right)
    -\frac{1}{2T}\left(V_x^2+V_y^2\right)\right.
\nonumber    \\
    &\qquad\left.
    +\,\frac{\Omega}{T}\left(
    \left(XV_y-YV_x\right)
    +\frac{b}{2}\left(X^2+Y^2\right)
    \right)
    \right],
\label{el}
\end{align}
where $\R = (X, Y, Z)$ and $\V = (V_x, V_y, V_z)$. Note that $\rho(Z, V_z)$ is not included in \eqref{el} because it factorizes out.

The rotating Gibbs distribution \eqref{el} is Gaussian, as expected. The non-zero moments are $\la X^2\ra = \la Y^2\ra$, $\la V_x^2\ra = \la V_y^2\ra$, and $\la X V_y\ra = -\la Y V_x\ra$. The three non-zero moments are determined from the following three equations:
\begin{align}
\label{pfm}
    T = \la V_x^2\ra-\frac{ \la X V_y\ra^2}{\la X^2\ra },
    \quad \Omega = \frac{ \la X V_y\ra}{\la X^2\ra }, \quad
    \frac{\la V_x^2 \ra}{\la X^2\ra}=\omega_0^2-b \Omega.
\end{align}
In the special case of zero magnetic field, we obtain:
\begin{subequations}
\label{generalized-gibbs-moments}
\begin{align}
\label{darwin1}
&    \la X^2\ra =\la Y^2\ra = \frac{T}{\omega_0^2-\Omega^2},
    \\
&    \la V_x^2\ra =\la V_y^2\ra = \frac{T\omega_0^2}{\omega_0^2-\Omega^2},
    \\
&    \la XV_y\ra = -\la YV_x\ra = \frac{T\Omega}{\omega_0^2-\Omega^2}.
\end{align}
\end{subequations}

\section{Vortical motion in fluid dynamics and stability condition (\ref{stability})}
\label{app-tornado}
Here we discuss generic rotation scenarios in fluid dynamics and relate them to the subject of section \ref{stabon}.

A cylindrically symmetric fluid (without artificial boundaries) cannot rotate as a rigid body, since its angular velocity $v_\phi=\Omega r$ (where $r$ and $\phi$ refer to cylindrical coordinates) will become unrealistically large for a constant $\Omega$ and a large $r$. For realistic vortices (e.g., tornados) at sufficiently large $r$, the rigid body rotation changes to the so-called potential vortex: $v_\phi\propto r^{-1}$ \cite{acheson}. The potential vortex is not realized for very small $r$, also because it predicts a very large $v_\phi$. The transition between two regimes is supported by several exactly solvable flow models, e.g., the Burgers vortex \cite{acheson}.

Such a transition between the two rotating regimes is also expected to occur for realistic rotating thermal baths, which are more similar to fluids than to rigid bodies. Let us now assume that the rotating Caldeira-Leggett oscillators also hold rigid-body rotation for small spatial extent (high frequencies) and vortex rotation $v_\phi\sim r^{-1}$ for larger spatial extent (smaller frequencies). The angular velocity $\Omega$ related to $v_\phi\sim r^{-1}$ scales as $\Omega\sim r^{-2}$. Estimating the equilibrium spatial extent $r_k$ of the $k$ low frequency oscillator via $r_k\sim\omega_k^{-1}$ [see e.g. (\ref{darwin1})], we see that the potential vortex rotation means $\Omega_k\sim \omega_k^2$ for a sufficiently small $\omega_k$.

We can now propose a specific model [cf.~(\ref{omega-tanh})]
\BEA
\Omega(w)=\Omega_0\tanh[\omega^2/\Omega_1^2],
\label{oshto}
\EEA
where $\Omega_1>0$ is a new characteristic frequency. According to (\ref{oshto}), $\Omega(w)\sim \omega^2$ for $\omega\ll \Omega_1$ (this models potential vortical motion), and $\Omega(\omega)\simeq \Omega_0$ for $\omega\gg \Omega_1$. 

\begin{figure}
    \centering
    \includegraphics[width=\linewidth]{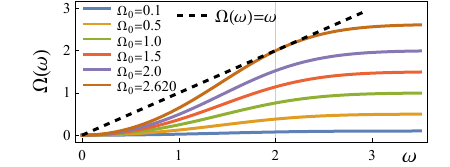}
    \includegraphics[width=\linewidth]{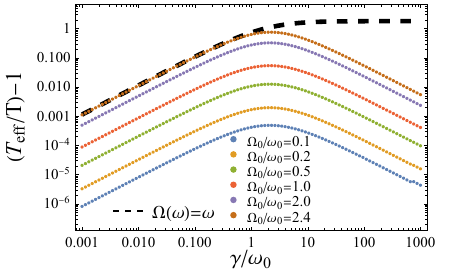}
    \includegraphics[width=\linewidth]{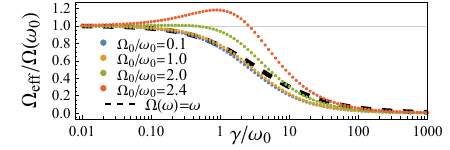}
    \caption{Top figure: rotational spectrum \eqref{oshto} for $\Omega_1=2$; cf.~Fig.~\ref{fig-omegas}. Middle figure: Effective temperature $T_{\rm eff}$ of the Brownian oscillator coupled to the rotating bath, depending on the friction coefficient $\gamma$; compare this with the left Fig.~(\ref{fig-effective-params}) and note that both figures have basically the same messages. Bottom figure: Effective angular velocity $\Omega_{\rm eff}$ {\it versus} $\gamma$. As compared to the right Fig.~(\ref{fig-effective-params}), the present bottom figure shows a new effect: $\Omega_{\rm eff}/\Omega(\omega_0)|_{\Omega_0/\omega_0=2.4}$ changes non-monotonically with $\gamma/\omega_0$, and there is a regime, where $\Omega_{\rm eff}/\Omega(\omega_0)|_{\Omega_0/\omega_0=2.4}>1$.   }
    \label{fig-vortex}
\end{figure}

In terms of (\ref{oshto}), condition (\ref{stability}) translates to a transcendental equation for $\Omega_0$ and $\Omega_1$. We shall not write it down in detail, but just note that (\ref{stability}) is certainly satisfied for $\Omega_1\geq \Omega_0$, and it is violated (at intermediate frequencies $\omega$) if $\Omega_1$ is sufficiently smaller than $\Omega_0$; a small, but positive $\Omega_0-\Omega_1$ still satisfies (\ref{stability}); see the top figure in Fig.~\ref{fig-vortex}.

Employing (\ref{oshto}) instead of (\ref{omega-tanh}) we get Fig.~\ref{fig-vortex} instead of Fig.~\ref{fig-effective-params}. Note that at the weak coupling limit $\gamma\to 0$, the rotating Gibbs distribution is recovered with parameters $T$ and $\Omega(\omega_0)$. We have $T_\eff >T$, i.e., the Brownian particle is hotter than the bath; see the middle Fig.~\ref{fig-vortex} and note that this conclusion agrees with (\ref{bala}). The bottom Fig.~\ref{fig-vortex} shows the scaled effective angular velocity $\Omega_\eff/\Omega(\omega_0)$. As compared to the right Fig.~(\ref{fig-effective-params}) and the first inequality in (\ref{ala}), the bottom Fig.~\ref{fig-vortex} shows a new effect: for a sufficiently large $\Omega_0/\omega_0$, $\Omega_{\rm eff}/\Omega(\omega_0)$ changes non-monotonically with $\gamma/\omega_0$. There is a regime, where $\Omega_{\rm eff}>\Omega(\omega_0)$. However, $\Omega_{\rm eff}<\Omega_0$, which agrees with the second inequality in (\ref{ala}).

\section{Cauchy Principal Value}
\label{app-cauchy}

The Cauchy Principal Value (CPV) is a technique used to assign finite values to certain improper integrals that would otherwise be undefined due to singularities. For instance, the integral of ${1}/{x}$ over the real line has a singularity at $x = 0$, and the CPV regularizes this by excluding an infinitesimal interval around $x = 0$. Specifically, the principal value of the integral is defined as the limit where a small symmetric region around the singularity is removed. Formally, the CPV of an integral is given by:
\begin{align}
    \calP \int_{-\infty}^\infty \frac{f(x)}{x} \d x &= \lim_{\epsilon \to 0} \left[ \int_{-\infty}^{-\epsilon} \frac{f(x)}{x} \d x + \int_{\epsilon}^\infty \frac{f(x)}{x} \d x \right]
    \nonumber
    \\
 &= \lim_{\epsilon \to 0}  \int_{\epsilon}^{\infty} \frac{f(x)-f(-x)}{x} \d x . 
 \label{Pval-def}
\end{align}
where $f(x)$ decays sufficiently fast at infinity. 
This definition ensures that integrals with singularities are well-defined by carefully handling the regions near $x = 0$ and isolating the ``principal'' part of the contribution.
For example,
\begin{align}
    \calP \int_{-1}^{1} \frac{1}{x} \, dx = \lim_{\epsilon \to 0} \left( \int_{-1}^{-\epsilon} \frac{1}{x} \, dx + \int_{\epsilon}^{1} \frac{1}{x} \, dx \right) = 0.
\end{align}
\comment{  
Another example with a higher-order pole is
\begin{equation}
    \calP \int_{-1}^{1} \frac{1}{x^2} \, dx = \lim_{\epsilon \to 0} \left( -\frac{1}{x} \Big|_{-1}^{-\epsilon} + -\frac{1}{x} \Big|_{\epsilon}^{1} \right) = -2.
\end{equation}
These integrals are divergent without the principal value.
}

One of the most important results involving the Cauchy Principal Value is the Sokhotski-Plemelj formula \cite{breuer2002theory}, which characterizes the behavior of the complex function $\frac{1}{x \pm i \varepsilon}$ as $\varepsilon \to 0^{+}$. The formula states that:
\begin{align}
\lim_{\varepsilon \to 0^{+}} \frac{1}{x \pm i \varepsilon} = \mathcal{P}\left(\frac{1}{x}\right) \mp i \pi \delta(x),
\label{sokho}
\end{align}
where $\delta(x)$ represents the Dirac delta function. Here, $\mathcal{P}\left(\frac{1}{x}\right)$ denotes the Cauchy Principal Value of $\frac{1}{x}$; see \eqref{Pval-def}. Neither $\delta(x)$ nor $\mathcal{P}\left(\frac{1}{x}\right)$ are ordinary functions; instead, they are distributions that act on test functions in integrals \cite{lighthill1958introduction}.
In the context of \eqref{sokho}, note \cite{lighthill1958introduction}:
\begin{align}
\label{dodosh}
\int\ee^{-i\omega t} \frac{\d\omega}{ 2\pi}= \delta(t),~~
\int\ee^{-i\omega t}\text{sgn}(\omega) \frac{i\,\d\omega}{2} = \calP\Big(\frac{1}{t}\Big).
\end{align}
This result is significant in various fields, including complex analysis, mathematical physics, and quantum field theory. The Cauchy Principal Value provides a method to handle singularities in integrals, while the delta function term captures the localized nature of the singularity at $x = 0$.

\comment{
 The Sokhotski–Plemelj formula is crucial in scenarios involving integrals of functions like $\frac{1}{x \pm i \varepsilon}$, as it decomposes the integral into real and singular components. In quantum field theory, this formula is often used in the calculation of propagators, where the infinitesimal imaginary term ensures that the Green's function respects causality.

The formula can be derived using contour integration techniques in the complex plane. By integrating $\frac{1}{x \pm i \varepsilon}$ around a contour that encircles the singularity at $x = 0$, it is shown that the result decomposes into a real part (the principal value) and an imaginary part (the delta function). This decomposition is also crucial in Fourier analysis, where it helps handle singularities in momentum space and regularize Fourier transforms.
}

\section{Evaluation of Integrals}
\label{app-inteval}

Here we evaluate the integral [cf.~\eqref{RKxi}]: 
\begin{align}
\label{XdX-int}
    \la \R\pp\dot\R\pp\tr{}\ra  \!= \!\!\int_0^t \!\! \int_0^t \! K(t\!-\!u)K'(t\!-\!u')\la \bxi\pp(u)\tr{\bxi\pp}(u')\ra\,\d u\d u'.
\end{align}
The components $\la X V_x\ra$ and $\la Y V_y\ra$ are straightforward to compute due to white noise: 
\BEA
\la \xi_x(t) \xi_x(t')\ra = \la \xi_y(t) \xi_y(t')\ra = \frac{2\gamma T}{1-a^2}\delta(t-t'), 
\EEA
as given in \eqref{noise-corr}. Therefore, we focus on the $\la X V_y\ra = - \la Y V_x\ra$ component:
\begin{align}
\label{intKs}
&    \la X V_y\ra = 
    \frac{2\gamma T \,a}{\pi(1-a^2)}
    \lim_{t\to\infty}
    \int_0^t\int_0^t K(u)K'(u')\times
    \nonumber\\
&    \calP\Big(\frac{1}{u-u'}\Big)
    \,\d u\d u',
\end{align}
where $K'(t)=\frac{\d }{\d t}K(t)$, and $K(t)$ is the inverse Laplace transform of $\hat K(s)$ from \eqref{K-hat}. The integral \eqref{intKs} can be rewritten as:
\begin{align}\label{intFs}
\la X V_y\ra =&\frac{2\gamma T \,a}{\pi(1-a^2)}\frac{1}{(\beta-\alpha)^2}\big\{
-\alpha F(\alpha,\alpha)
+\beta F(\alpha,\beta)
\nonumber
\\
&\qquad+\alpha F(\beta,\alpha)
-\beta F(\beta,\beta)
\big\}
\end{align}
where the function $F$ is defined as:
\begin{align}
\label{xyint}
    F(\mu,\nu)\equiv\int_0^\infty\int_0^\infty \ee^{-\mu x}\ee^{-\nu y}
    \;
    \calP\Big(\frac{1}{x-y}\Big) 
    \,\d x\d y
\end{align}
for positive $\mu$ and $\nu$. Applying the change of variables $t = x+y$ and $s = x-y$, \eqref{xyint} becomes:
\begin{align}
    =&\frac{1}{2}\int_0^\infty\!\!\d t\,\ee^{-\frac{1}{2}(\mu+\nu)t} \int_{-t}^{t}\d s \,\ee^{-\frac{1}{2}(\mu-\nu)s}\;\calP\Big(\frac{1}{s}\Big)
\\
=&
    \frac{1}{2}\int_0^\infty\!\!\d t\,\ee^{-\frac{1}{2}(\mu+\nu)t} \int_{0}^{t}\d s \,\frac{1}{s}\left[\ee^{-\frac{1}{2}(\mu-\nu)s}-\ee^{+\frac{1}{2}(\mu-\nu)s}\right]
\end{align}
where we used \eqref{Pval-def} for the principal value. Changing the order of integration:
\begin{align}
    =&\frac{1}{2}\int_0^\infty\!\!
    \d s\;
    \frac{1}{s}\left[\ee^{-\frac{1}{2}(\mu-\nu)s}-\ee^{+\frac{1}{2}(\mu-\nu)s}\right]
    \int_s^\infty\!\! \d t\,\ee^{-\frac{1}{2}(\mu+\nu)t}
    \nonumber
    \\
    =&
    \frac{1}{\mu+\nu}\int_0^\infty\!\!
    \d s\;
    \frac{1}{s}\left[\ee^{-\frac{1}{2}(\mu-\nu)s}-\ee^{+\frac{1}{2}(\mu-\nu)s}\right]\ee^{-\frac{1}{2}(\mu+\nu)s}
    \nonumber
    \\
    =& \frac{1}{\mu+\nu}\int_0^\infty
    \d s\;
    \frac{1}{s}\left[\ee^{-\mu s}-\ee^{-\nu s}\right]
    \\
    =&\frac{1}{\mu+\nu}\int_0^\infty
    \d s\;
    \frac{1}{s}\,\ee^{-\frac{\mu}{\nu+\mu}s}\left(1-\ee^{-s}\right)
    \\
    =&\frac{1}{\nu+\mu}\log\left(\frac{\nu}{\mu}\right) \equiv F(\mu,\nu).
    \label{F-value}
\end{align}
where in the final line we used
\begin{align}
    I(p)=\int_0^\infty\d s\;
    \frac{1}{s}\,\ee^{-p s}\left(1-\ee^{-s}\right) = \log\left(\frac{1+p}{p}\right)
\end{align}
for $p>0$, which can be shown by noting that $\frac{\d}{\d p}I = \frac{1}{1+p}-\frac{1}{p}$ and $I(\infty) = 0$.
Substituting \eqref{F-value} into \eqref{intFs}, we obtain:
\begin{align}
    \la X V_y\ra =& 
    \frac{2\gamma T}{\pi}
    \frac{a}{1-a^2}\frac{\log (\alpha )-\log (\beta )}{\alpha^2-\beta^2}.
\end{align}
which is equivalent to \eqref{xvy-line3}.

\section{\WK{} theorem}
\label{app-wiener}

Consider two stationary processes $A(t)$ and $B(t)$. Define their correlation functions as follows:
\begin{align}
    \rC_{AB}(t'-t)=&\la A(t')B(t)\ra
    \\
    =& \rC_{BA}(t-t')
    \\
    \rC_{AB}(t)=&\rC_{BA}(-t).
\end{align}
Let $\td{A}(\omega)$, $\td{B}(\omega)$, and $\rS_{AB}(\omega)$ denote the Fourier transforms of $A(t)$, $B(t)$, and $\rC_{AB}(t)$, respectively. The function $\rS$ is known as the spectral density. The Wiener-Khinchin theorem states:
\begin{align}
    \la \td{A}(\omega')\td{B}(\omega)\ra =& 2\pi\,\delta(\omega'+\omega)\,\rS_{AB}(\omega')
    \\
    =& 2\pi\,\delta(\omega'+\omega)\,\rS_{AB}(-\omega)
    \\
    =& 2\pi\,\delta(\omega'+\omega)\,\rS_{BA}(\omega).
\end{align}
The proof of this theorem is straightforward:
\begin{align}
    &\la \td{A}(\omega')\td{B}(\omega)\ra = \int_{-\infty}^\infty \int_{-\infty}^\infty \!\!\la A(t')B(t) \ra \ee^{-\ii\omega' t'}\ee^{-\ii\omega t}\,\d t\,\d t'
    \nonumber
        \\
    &\qquad= \int_{-\infty}^\infty \left[\int_{-\infty}^\infty \!\!\!\rC_{AB}(t'-t) \ee^{-\ii\omega'(t'-t)} \d t' \right]
    \ee^{-\ii (\omega+\omega')t} \d t
    \nonumber
        \\
    &\qquad=  \rS_{AB}(\omega') \int_{-\infty}^\infty \ee^{-\ii (\omega+\omega')t} \d t
        \\
    &\qquad= \rS_{AB}(\omega') \;2\pi \,\delta(\omega'+\omega).
\end{align}

\section{Stationary state in a magnetic field}
\label{app-magnetic}

In the presence of a magnetic field ($b\ne 0$), the integrals in \eqref{RR} and \eqref{dotRR} take the following form:
\begin{subequations}
\label{integrals-b-app}
\begin{align}
\label{integrals-b-1-app}
    \la X^2 \ra = \la Y^2 \ra =& \int_0^\infty {\phi_1(\omega)}{g(\omega)}\;\d\omega, \\
    \label{integrals-b-2-app}
    \la X V_y\ra = -\la Y V_x \ra =& \int_0^\infty {\phi_2(\omega)}{g(\omega)}\;\d\omega, \\
    \label{integrals-b-3-app}
    \la V_x^2 \ra = \la V_y^2 \ra =& \int_0^\infty {\phi_3(\omega)}{g(\omega)}\;\d\omega.
\end{align}
\end{subequations}
where 
\begin{align}
&    \phi_1(\omega) = \frac{1}{\pi}\frac{2 \gamma T \omega^2}{\omega^2 - \Omega(\omega)^2}
\times\\
&\qquad \times\left[\omega^2 \left(b^2+\gamma^2\right)+\left(\omega^2-\omega_0^2\right) \left(\omega^2-\omega_0^2-2 b \Omega (\omega )\right)\right], \nonumber
\\
&    \phi_2(\omega) = \frac{1}{\pi}\frac{2 \gamma T \omega^2}{\omega^2 - \Omega(\omega)^2}
 \Big[ \omega^2 \left(b^2+\gamma^2\right) \Omega (\omega )  \\
&\qquad\qquad\qquad+\left(\omega^2-\omega_0^2\right) \left(\Omega (\omega ) \left(\omega^2-\omega_0^2\right)-2 b \omega^2\right)\Big], \nonumber
\\
&    \phi_3(\omega) =\frac{1}{\pi}\frac{2 \gamma T \omega^4}{\omega^2 - \Omega(\omega)^2}\times  \\
&\qquad\times\Big[\omega^2 \left(b^2+\gamma^2\right) 
+\left(\omega^2-\omega_0^2\right) \left(\omega^2-\omega_0^2-2 b \Omega (\omega )\right)\Big],\nonumber
\\
& g(\omega)=\left[\prod_{i=1}^{4} (\omega-\omega_i)(\omega-\omega^\ast_i) \right]^{-1},
\end{align}
where in $g(\omega)$ 
\begin{subequations}
    \label{poles-def}
\begin{align}
    \omega_1 =& \frac{1}{2}\left(\hat\beta + \sqrt{\hat\beta^2+4\omega_0^2}\right) \quad\text{with}\quad \hat\beta= b + \ii \gamma, \\
    \omega_2 =& \frac{1}{2}\left(\hat\beta + \sqrt{\hat\beta^2+4\omega_0^2}\right) \quad\text{with}\quad \hat\beta= -b + \ii \gamma, \\
    \omega_3 =& \frac{1}{2}\left(\hat\beta - \sqrt{\hat\beta^2+4\omega_0^2}\right) \quad\text{with}\quad \hat\beta= b + \ii \gamma, \\
    \omega_4 =& \frac{1}{2}\left(\hat\beta - \sqrt{\hat\beta^2+4\omega_0^2}\right) \quad\text{with}\quad \hat\beta= -b + \ii \gamma.
\end{align}
\end{subequations}
These roots are illustrated in Figure \ref{fig-poles}. As $\gamma \to 0$, these roots approach the real axis in the complex plane, causing $g(\omega)$ to exhibit peaks at these locations. For positive $\omega$, only $\omega_1$, $\omega_2$, and their complex conjugates significantly affect the behavior of $g(\omega)$:
\begin{align}
    g(\omega) = g_\text{slow}(\omega)g_\text{fast}(\omega), \\
    g_\text{fast}(\omega) = \frac{1}{|\omega-\omega_1|^2|\omega-\omega_2|^2}.
\end{align}
Furthermore, if the real parts of these roots are well separated, we may approximate
\begin{align}
&    g_\text{fast}(\omega) \approx 
    \frac{1}{|\omega-\omega_1|^2|\omega_1-\omega_2|^2}
    +
    \frac{1}{|\omega_2-\omega_1|^2|\omega-\omega_2|^2}
\nonumber    \\
&    \approx 
    \frac{\pi}{\omega''_1}\delta\left(\omega-\omega'_1\right)\frac{1}{|\omega_1-\omega_2|^2}
    +
    \frac{1}{|\omega_2-\omega_1|^2} 
    \frac{\pi}{\omega''_2}
    \delta\left(\omega-\omega'_2\right),
\end{align}
where in the second line, $\omega'_i$ ($\omega''_i$) is the real (imaginary) part of root $\omega_i$, and we used the following limit of the Dirac delta function:
\begin{align}
\label{delta-limit}
    \frac{\epsilon}{\pi (x^2+\epsilon^2)} \rightarrow \delta(x), \quad\text{when}\quad
    \epsilon \rightarrow 0.
\end{align}
Now, the moments just evaluate into
\begin{align}
    \phi_i(\omega'_1)g_\text{slow}(\omega'_1)\frac{\pi}{\omega''_1}\frac{1}{|\omega_1-\omega_2|^2} \,+\qquad\nonumber\\
    + \,\phi_i(\omega'_2)g_\text{slow}(\omega'_2)\frac{\pi}{\omega''_2}\frac{1}{|\omega_1-\omega_2|^2}.
\end{align}
Using these expressions of the moments, next, using \eqref{pfm}, we evaluate the effective temperature and the effective angular velocity of the Gibbs distribution:
\begin{align}
    \Teff =& T \frac{\omega_+ + \omega_-}{\omega_+ + \omega_- + \Omega(\omega_+)-\Omega(\omega_-)}, \\
    \Oeff =& \frac{\omega_+ \Omega(\omega_-)+ \omega_- \Omega(\omega_+)}{\omega_+ + \omega_- + \Omega(\omega_+)-\Omega(\omega_-)},
\end{align}
where $\omega_+$ ($\omega_-$) are $\omega_1$ ($\omega_2$) at the $\gamma \to 0$ limit:
\begin{align}
\omega_+=\frac{\sqrt{4\omega_0^2 + b^2}+b}{2}, \quad
\omega_- = \frac{\sqrt{4\omega_0^2 + b^2}-b}{2}.
\end{align}

\section{Vector Potential in Coulomb Gauge for Circular Current Loop}
\label{app-loop}

The purpose of this Appendix is to evaluate the vector potential $\A(\R)$ in the Coulomb gauge ${\rm div}\A=0$ for a magnetic field $\B(\R)$ which goes to zero for $|\R|\to\infty$, but is 
approximately homogeneous for $|\R|\sim 0$. 

To this end, we recall
that cylindrical coodinates $r$, $\phi$ and $z$ are defined via the
rectangular coordinates $X$, $Y$ and $Z$: $X=r\cos \phi$ and $Y=r\sin
\phi$. The correspoding unit vectors are 
\begin{align}
&  \bm{e}_r=\cos\phi\, \bm{e}_x+\sin\phi \,\bm{e}_y ,\qquad 
  \bm{e}_\phi=-\sin\phi \,\bm{e}_x+\cos\phi\, \bm{e}_y,\nonumber\\
&  \bm{e}_r\times \bm{e}_\phi=\bm{e}_z,\quad
\bm{e}_z\times \bm{e}_r=\bm{e}_\phi,\quad 
\bm{e}_\phi\times \bm{e}_z=\bm{e}_r.
  \label{eq:54}
\end{align}
Any vector expands as 
\begin{eqnarray}
  \label{eq:55}
  \A=  A_r\bm{e}_r+  A_{\phi}\bm{e}_\phi +A_z \bm{e}_z.
\end{eqnarray}
Next, let us recall that $\A(\R)$ in the Coulomb gauge directly relates to the electric current $\bm{J}(\R)$, which generates the magnetic field \cite{landau2} [we take $c=1$]:
\BEA
\label{ugar}
\A(\R)=\int\d^3 R'\,\frac{\bm{J}(\R')}{|\R-\R'|}.
\EEA
We consider $\bm{J}$ that within (\ref{eq:55}) has only non-zero component $J_\phi(r,z)$, which does not depend on $\phi$. The conservation condition ${\rm div}\bm{J} = 0$ is automatically satisfied since:
\BEA
{\rm div}\bm{J}
= \frac{1}{\rho} \frac{\partial(\rho J_\rho)}{\partial \rho} + \frac{1}{\rho} \frac{\partial J_\phi}{\partial \phi} + \frac{\partial J_z}{\partial z} = 0.
\EEA
Using (\ref{eq:54}, \ref{eq:55}) we transform \eqref{ugar} to cylindrical coordinates and find
\begin{align}
\label{ugar2}
&A_x(\R)=-\int_0^\infty\int_0^{2\pi}\int r\d r\,\d\phi\, \d z \nonumber\\
& \qquad \frac{{J}_\phi(r,z)\sin\phi }{\sqrt{(X-r\cos\phi)^2
+(Y-r\sin\phi)^2+(Z-z)^2}}, \\
&A_y(\R)=\int_0^\infty\int_0^{2\pi}\int r\d r\,\d\phi\, \d z \nonumber\\
&\qquad\frac{{J}_\phi(r,z)\cos\phi }{\sqrt{(X-r\cos\phi)^2
+(Y-r\sin\phi)^2+(Z-z)^2}},
\label{ugar3}
\end{align}
where $\R=(X,Y,Z)$ and $A_z=0$. The simplest way of evaluating (\ref{ugar2}, \ref{ugar3}) is to assume that $J_\phi(r,z)$ as a function of $r$ and $z$ is localized:
\BEA
J_\phi(r,z)=\delta(r-R_0) \delta(z )\hat J_\phi,
\EEA
and then look at (\ref{ugar2}, \ref{ugar3}) in the domain $|\R|\ll R_0$:
\BEA
\A=\frac{\hat J_\phi\pi }{2R_0}[-Y,X,0]+{\cal O}(|\R|^2/R_0^2),
\EEA
which agrees with (\ref{gauge-choice}).

\section{Existence of free energy}
\label{app-proof}
In this section, we address the existence of the free energy for a system described by the Langevin equation \eqref{lang}. Specifically, we show that if the Hamiltonian of the system depends on parameters $\lambda_1, \lambda_2, \ldots$, then the slow work (\ref{W-def3})
is a complete differential. We demonstrate this for rotation symmetric systems coupled to the rotational thermal bath, focusing on the case where the Hamiltonian is quadratic in the canonic momentum $\P$ and canonic position $\R$.

Given that isothermal work is defined in terms of the Hamiltonian (see \eqref{W-def2}), we express the equations of motion using the system Hamiltonian $H_S(\R, \P)$, where $\R$ and $\P$ represent the canonic positions and momenta, respectively. Following the procedure of Section \ref{sec-cl}, the Langevin equation for 
in coordinate and canonical momentum \eqref{ghana} reads:
\begin{subequations}
\label{H-lang}
\begin{align}
    \dot{\P}=&-\pd{H_S}{\R} -\int_0^t\d u\,\zeta(t-u)\dot{\R}(u)+\bxi,
    \\
    \label{H-lang-line2}
    \dot{\R}=&+\pd{H_S}{\P}.
\end{align}
\end{subequations}
Here, only one of Hamilton's equations is modified to include noise and friction. The friction term involves the velocity $\dot{\R}$ rather than the canonic momentum $\P$. The definitions of the memory kernel $\zeta(t)$ and the noise spectrum $\xi(t)$ remain the same as in \eqref{C-gen} and \eqref{zeta-gen}. In the special case where $H_S$ corresponds to the Lagrangian \eqref{lagr_sys}, the system of first-order stochastic differential equations \eqref{H-lang} precisely matches the second-order Langevin equation \eqref{lang}.

\comment{The second Hamilton's equation \eqref{H-lang-line2} indicates that the canonic momentum $\P$ and the kinematic momentum $\dot{\R}$ are generally not equal. Specifically, for the system defined by the Lagrangian \eqref{lagr_sys}, the canonic momentum is given by
\begin{align}
    \P = \dot{\R} + A(\R) = \dot{\R} + \frac{1}{2} \B \times \R,
\end{align}
where we have used the Coulomb gauge \eqref{gauge-choice} ${\rm div} \A = 0$. When changes to the system are slow, the Coulomb gauge and the Lorenz gauge become identical. \TODO{<-Check this}
}

Rotating thermal bath imposes strict constraints on the relationship between memory kernel $\zeta(t)$ and noise spectrum $\rS_\bxi(\omega)$, see \eqref{S} and \eqref{zeta-gen}. However, here we consider a general memory kernel $\zeta(t)$ and noise spectrum $\rS_\bxi(\omega)$.
Furthermore, we assume that the Hamiltonian $H_S$ is invariant under rotations around a fixed axis. This makes the overall system rotation symmetric. The axis of symmetry can be defined, for instance, by a uniform magnetic field or a confining potential with cylindrical symmetry.

As discussed above, we restrict our analysis to the quadratic form of the Hamiltonian:
\begin{align}
    \label{hashes}
    H_S = \frac{1}{2} \begin{pmatrix}
        \tr{\R} ~ \tr{\P}
    \end{pmatrix}
    \begin{pmatrix}
        \Sigma_{RR} & \Sigma_{RP}
        \\
        \tr{\Sigma_{RP}} & \Sigma_{PP}
    \end{pmatrix}
    \begin{pmatrix}
        {\R} \\ {\P}
    \end{pmatrix},
    \\
    \Sigma = \begin{pmatrix}
        \Sigma_{RR} & \Sigma_{RP}
        \\
        \tr{\Sigma_{RP}} & \Sigma_{PP}
    \end{pmatrix},
\end{align}
where $\Sigma$ is a symmetric $2d \times 2d$ matrix, with $d=3$ for three-dimensional systems. If the system exhibits rotation symmetry around the $z$-axis, the most general form of the Hamiltonian can be parameterized as:
\begin{equation}
    \label{gen-sigma}
\begin{gathered}
    \Sigma_{RR} = \begin{pmatrix}
       \frac{b^2 }{4m}+ m \omega_0^2 & 0 & 0
        \\
        0 & \frac{b^2}{4m} + m \omega_0^2 & 0
        \\
        0& 0 & m_z \omega_{z0}^2
    \end{pmatrix},
    \\
    \Sigma_{PP} = \begin{pmatrix}
        \frac{1}{m}  & 0 & 0
        \\
        0 & \frac{1}{m}  & 0
        \\
        0& 0 & \frac{1}{m_z}
    \end{pmatrix},
    \quad
    \Sigma_{RP} = \begin{pmatrix}
        c & -\frac{b }{2m} & 0
        \\
        +\frac{b }{2m} & c & 0
        \\
        0 & 0 & c_z
    \end{pmatrix}.
\end{gathered}
\end{equation}
Thus, the most general rotation symmetric quadratic Hamiltonian has seven independent parameters $\lambda_\alpha$. In the specific case where $c = c_z = 0$, $m_z = m = 1$, and $b$ represents the magnitude of a uniform magnetic field along the $z$-axis, the Hamiltonian \eqref{hashes} corresponds to the Lagrangian \eqref{lagr_sys} with a harmonic, rotation symmetric external potential $V(\R) = \frac{\omega_0^2}{2}(X^2 + Y^2) + \frac{\omega_{0z}^2}{2}Z^2$ and the gauge \eqref{gauge-choice}.

\comment{
Recall the definition of the noise spectrum $\rS_\bxi(\omega)$:
\begin{align}
    \la \tilde{\bxi}(\omega)\tilde{\bxi}\tr{}(\omega') \ra = 2\pi \delta(\omega+\omega') \rS_\bxi(\omega).
\end{align}
For our thermal bath, the noise exhibits a specific spectrum given by \eqref{S}, determined by the thermal bath parameters. However, we now consider its most general form.
By definition, $\rS_\bxi(\omega) = \rS_\bxi(\omega)^\dagger = \rS_\bxi(-\omega)\tr{}$. The rotational symmetry of the overall system imposes additional constraints on the form of the spectrum:
\begin{align}
\label{xi-spectr}
    \rS_\bxi(\omega) = \begin{pmatrix}
        S_1(\omega) & \ii S_2(\omega) & 0
        \\
        - \ii S_2(\omega) & S_1(\omega) & 0
        \\
        0 & 0 & S_z(\omega)
    \end{pmatrix},
\end{align}
where $S_z$ and $S_1$ are real even functions, and $S_2$ is a real odd function. However, these properties will not be directly relevant to the subsequent derivation.
}

We now determine the stationary state distribution of the system, following the approach outlined in Section \ref{sec-ft}, which will be utilized in \eqref{W-def3}. Let $\rS(\omega)$ represent the $2d \times 2d$ spectral density of the row vector $\left(\R(t)\tr{},\, \P(t)\tr{} \right)$. The second moment is given by
\begin{align}\label{RP-moment}
     \la \begin{pmatrix}
        \tr{\R} ~ \tr{\P}
    \end{pmatrix}
    \begin{pmatrix}
        {\R} \\ {\P}
    \end{pmatrix} \ra = \rC(0) = \int_{-\infty}^\infty \rS(\omega)\,\d\omega,
\end{align}
where $\rC(t)$ denotes the $6 \times 6$ autocorrelation function. The averages can be expressed in terms of $\rS(\omega)$ as follows:
\begin{align}
    \label{avg-tr}
    \la \pd{H_S}{\lambda_\alpha} \ra 
    = \int\d\omega \Tr\left(\rS \pd{\Sigma}{\lambda_\alpha}\right),
\end{align}
where $\Tr$ denotes the trace of the matrix.
Note that only $\Sigma$ in \eqref{hashes} depends on $\lambda_\alpha$. 

To establish a relationship between $\rS$ and $\rS_\bxi$, consider the Fourier transform of \eqref{H-lang}:
\begin{subequations}
\label{H-eqns}
\begin{align}
    \ii\omega \tilde{\P}=&-\pd{H_S}{\R}(\tilde{\R},\tilde{\P})
    -\ii\omega\tilde\zeta(\omega)\tilde{\R}+\tilde\bxi,
    \label{H-eqns-l1}
    \\
    \ii\omega\tilde{\R}=&+\pd{H_S}{\P}(\tilde{\R},\tilde{\P}),
    \label{H-eqns-l2}
\end{align}
\end{subequations}
Solving these equations for $\tilde{\R}$ and $\tilde{\P}$, we obtain:
\begin{align}
\label{H-lang-ft}
    &\begin{pmatrix}
        \tilde{\R} \\ \tilde{\P}
    \end{pmatrix} = \rM \begin{pmatrix}
        \tilde{\bxi} \\ 0
    \end{pmatrix},
    \\
    &\text{with}\quad
    \rM(\omega)^{-1} = \begin{pmatrix}
        \ii\omega \tilde\zeta \rI & \ii\omega \rI
        \\
        -\ii\omega \rI & 0
    \end{pmatrix} + \Sigma \equiv \rK(\omega)
    \label{M-def}
\end{align}
Note that $\rM^{-1}$ is linear in $\Sigma$, such that $\d(\rM^{-1}) = \d \,\Sigma$. 

Since $\zeta(t)$ is real, its Fourier transform satisfies the conjugate symmetry condition: $\tilde{\zeta}(-\omega) = \tilde{\zeta}(\omega)^\ast$. Consequently, $\rM(-\omega) = \rM(\omega)^\dagger$. The spectral density can then be expressed as:
\begin{align}
\label{SM}
    &\rS(\omega)  =  \rM(\omega)
    \bar\rS_\bxi(\omega)
    \rM(\omega)^\dagger,
    \\
    &
    \text{with}\quad
    \bar\rS_\bxi(\omega) = \begin{pmatrix}
        \rS_\bxi(\omega) & 0
        \\
        0 & 0
    \end{pmatrix}.
    \label{Sxi-def}
\end{align}

Substituting this into \eqref{avg-tr}, we obtain
\begin{align}
    \frac{\delta W}{\delta \lambda_\alpha} =& \la \pd{H_S}{\lambda_\alpha} \ra 
    = \int\d\omega \Tr\left(\rM \bar\rS_\bxi \rM^\dagger \pd{\rM^{-1}}{\lambda_\alpha}\right),
\end{align}
where we utilized the property $\pd{}{\lambda} \rM^{-1} = \pd{}{\lambda} \Sigma$, as seen in \eqref{M-def}. Recall the identity for the differential of the inverse matrix:
\begin{align}
    \label{invdiff}
    \tfrac{\d}{\d \lambda} \rM^{-1} = - \rM^{-1}\left(\tfrac{\d}{\d \lambda} \rM\right) \rM^{-1}.
\end{align}
Using this, we can rewrite the expression as
\begin{align}
    \frac{\delta W}{\delta \lambda_\alpha} =& -\int\d\omega \Tr\left(\rM \bar\rS_\bxi \rM^\dagger \rM^{-1}\pd{\rM}{\lambda_\alpha}\rM^{-1}\right)
    \\
    =& -\int\d\omega \Tr\left(\bar\rS_\bxi \rM^\dagger \rM^{-1}\pd{\rM}{\lambda_\alpha}\right)
    \\
    =& -\int\d\omega \sum_{i,j=1}^d \rS^\xi_{ij}
    \left[ \rM^\dagger \rM^{-1}\pd{\rM}{\lambda_\alpha}\right]_{ij},
    \label{eachsummand}
\end{align}
where in the second line we applied the cyclic property of the trace. Note that $\rS_\bxi$ corresponds to the non-zero upper left block of $\bar\rS_\bxi$, as defined in \eqref{Sxi-def}.

We claim that the term within the square brackets forms a complete differential for any $1 \le i, j \le d$:
\begin{align}
    \label{claim1}
    \d f = \sum_\alpha \left[\rM^\dagger \rM^{-1}\pd{\rM}{\lambda_\alpha}\right]_{ij}\,\d\lambda_\alpha
\end{align}
for some function $f(\lambda_1, \lambda_2, \dots; \omega)$. This claim implies the existence of a free energy $F(\lambda_1, \lambda_2, \dots)$ such that $\frac{\delta W}{\delta \lambda_\alpha} = \pd{F}{\lambda_\alpha}$. We now proceed to prove \eqref{claim1}.

For convenience, define the following $2d\times2d$ block matrix:
\begin{align}
    \rU = \begin{pmatrix}
        \rI & 0
        \\
        0 & 0
    \end{pmatrix}.
\end{align}
Equivalently, we need to show that the following $2d\times2d$ matrix is a complete differential:
\begin{align}
    \label{tosimpl}
    \sum_\alpha\rU\rM^\dagger \rM^{-1}\pd{\rM}{\lambda_\alpha} \rU \,\d\lambda_\alpha
\end{align}
We  further simplify \eqref{tosimpl} using the relation 
\begin{align}
\rM^{-1} = \left(\rM^\dagger\right)^{-1} + \ii\omega \left(\td{\zeta}+\td{\zeta}^\ast\right) \rU
\end{align}
which follows from the definition \eqref{M-def} and the property $\left(\rM^\dagger\right)^{-1}=\left(\rM^{-1}\right)^\dagger$.
We substitute this into \eqref{tosimpl} to obtain:
\begin{align}
    \rU\rM^\dagger \rM^{-1}\pd{\rM}{\lambda} \rU =& 
    \rU\left(\rI + \rM^\dagger \ii\omega \left(\td{\zeta}+\td{\zeta}^\ast\right) \rU \right)\pd{\rM}{\lambda} \rU
    \nonumber
    \\
    =&\rU\pd{\rM}{\lambda} \rU + 
    \rU\rM^\dagger\rU \pd{}{\lambda}\left(\rU \rM \rU \right),
    \label{remains}
\end{align}
where in the last line we used the identity $\rU^2=\rU$. Note that the first term is already a complete differential, $\rU\pd{\rM}{\lambda} \rU\d\lambda_\alpha = \rU(\d\rM) \rU$, so it remains to show that the second term is also a complete differential. First, note that $\rU \rM \rU \equiv \rN$ is the upper left block of $\rM$. From \eqref{H-lang-ft}, the matrix $\rN$ relates the position and noise
\begin{align}
    \label{Rxi}
    \td{\R} = \rN \td{\bxi}.
\end{align}
By substituting $\td{\P}$ from \eqref{H-eqns-l2} into \eqref{H-eqns-l1}, we find the expression for $\rN$:
\begin{align}
    \rN^{-1} =& \left[ 
    \left(\ii\omega \rI + \Sigma_{RP}\right)
    \Sigma_{PP}^{-1}
    \left(\ii\omega \rI - \tr{\Sigma_{RP}}\right)
    + \Sigma_{RR}
    + \ii \omega \zeta \rI
    \right]
    \nonumber
    \\
    =&\left[ 
    \rH
    + \ii \omega \zeta \rI
    \right],
    \label{ninv2}
\end{align}
where $\rH = \rH^\dagger=\left(\ii\omega \rI + \Sigma_{RP}\right)
    \Sigma_{PP}^{-1}
    \left(\ii\omega \rI - \tr{\Sigma_{RP}}\right)
    + \Sigma_{RR}$ is a Hermitian matrix. Note that \eqref{Rxi} and the rotation symmetry of the overall system imply that the matrix $\rN$ is invariant under rotations around the symmetry axis.

In \eqref{remains} remains to show that $\rN^\dagger\pd{}{\lambda_\alpha}\rN \,\d\lambda_\alpha$ is a complete differential. Here we will employ another approach: we will show that the mixed derivatives are equal:
\begin{align}
    0 =&\; \pd{}{\lambda_\beta}\left(\rN^\dagger\pd{\rN}{\lambda_\alpha}\right) -
    \pd{}{\lambda_\alpha}\left(\rN^\dagger\pd{\rN}{\lambda_\beta}\right),
    \\
    =&\;\pd{\rN^\dagger}{\lambda_\beta}\pd{\rN}{\lambda_\alpha} -
    \pd{\rN^\dagger}{\lambda_\alpha}\pd{\rN}{\lambda_\beta}.
    \label{toshow}
\end{align}
Applying the inverse formula \eqref{invdiff} to the derivative of $\rN$, we obtain:
\begin{align}
    \pd{\rN}{\lambda_\alpha} =& \pd{}{\lambda_\alpha} \left[ 
        \rH
        + \ii \omega \zeta \rI
        \right]^{-1}
        = -\rN \pd{\rH}{\lambda_\alpha} \rN
        \\
    \pd{\rN^\dagger}{\lambda_\alpha} =& -\rN^\dagger \pd{\rH}{\lambda_\alpha} \rN^\dagger
\end{align}
where we used the fact that only $\rH$ in \eqref{ninv2} depends on the system parameters and $\rH$ is Hermitian. Substituting these into \eqref{toshow}, we get:
\begin{align}
    \label{allcommute}
    \rN^\dagger \pd{\rH}{\lambda_\alpha} \rN^\dagger \rN \pd{\rH}{\lambda_\beta} \rN - \rN^\dagger \pd{\rH}{\lambda_\beta} \rN^\dagger \rN \pd{\rH}{\lambda_\alpha} \rN.
\end{align}
Up to this point, we have not utilized the rotation symmetry of the overall system. All matrices in \eqref{allcommute} are invariant under rotation around a fixed axis. In $d=3$ dimensions, such rotations are represented by a tensor product of a 2D rotation and the identity matrix. Since 2D rotations commute, the same property holds for the tensor product. Therefore, $\rN$, $\rN^\dagger$, $\pd{\rH}{\lambda_\alpha}$, and $\pd{\rH}{\lambda_\beta}$ commute with each other. Consider the first term in \eqref{allcommute}:
\begin{align}
    \rN^\dagger \pd{\rH}{\lambda_\alpha} \rN^\dagger \rN \pd{\rH}{\lambda_\beta} \rN 
    =& \rN^\dagger  \rN^\dagger \rN \pd{\rH}{\lambda_\beta} \pd{\rH}{\lambda_\alpha}\rN
    \\=& 
    \rN^\dagger  \pd{\rH}{\lambda_\beta}\rN^\dagger \rN  \pd{\rH}{\lambda_\alpha} \rN.
\end{align}
This makes \eqref{allcommute} zero, thus proving \eqref{toshow}. Consequently, we have shown that $\rN^\dagger \pd{}{\lambda_\alpha} \rN \, \d\lambda_\alpha$ is a complete differential, as well as \eqref{remains}, and finally \eqref{W-def3}.

Importantly, this result is independent of the specific form of the noise spectrum \eqref{Sxi-def}. We have demonstrated that each summand in \eqref{eachsummand} is a complete differential, which depends solely on the form of the dynamical equations \eqref{H-lang} and the rotation symmetry of the system.

\section{Non-rotation-symmetric potential}
\label{app-aniso}
\begin{figure}
    \centering
    \begin{minipage}{0.48\linewidth}
        \includegraphics[width=\linewidth]{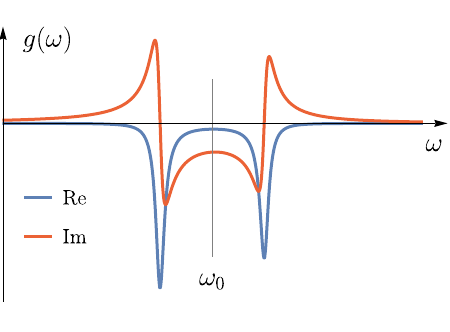}   
    \end{minipage}\hfill%
    \begin{minipage}{0.48\linewidth}
            \includegraphics[width=\linewidth]{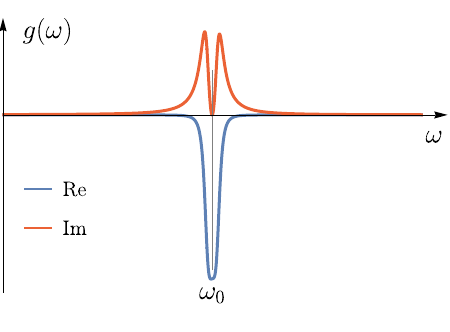} 
    \end{minipage}
    
    \caption{
    \textbf{The function $g(\omega)$ defined in \eqref{g-def} exhibits peaks around $\omega_0$. The peak width is determined by the parameter $\gamma$.}
    \textbf{(left)} When $\gamma/\omega_0 \ll \epsilon \ll 1$, the peaks are well-separated and located at $\omega_0(1\pm\epsilon)$. As $\gamma \to 0$, these peaks become increasingly sharp.
    \textbf{(right)} When $\epsilon \ll \gamma/\omega_0 \ll 1$, the peak width exceeds the separation between the peaks, resulting in a single broad peak around $\omega_0$.
    }
    \label{fig-gplot}
\end{figure}

In this section, we evaluate the integral \eqref{XY-int} in the limits $\gamma \to 0$ and $\epsilon \to 0$:
\begin{align}
\label{XY-gphi}
&\la X Y\ra = \Re \int_{0}^\infty
g(\omega)\phi(\omega)\,\d\omega,
\\
&g(\omega) = \frac{\ii}{
    (\omega_x^2-\omega^2+\ii\gamma\omega)
    (\omega_y^2-\omega^2-\ii\gamma\omega)},
\\\label{g-def}
&\phi(\omega) = \frac{2T\gamma\omega \Omega(\omega)}{\pi(\omega^2-\Omega(\omega)^2)}.
\end{align}
The function $g(\omega)$ is illustrated in Figure \ref{fig-gplot}.

The limit is not unique and depends on the order of taking the limits $\gamma \to 0$ and $\epsilon \to 0$.
In the following subsections, we consider these cases separately.

\subsection{$\epsilon \ll \frac{\gamma}{\omega_0} \ll 1$}
In this case, $\omega_x \approx \omega_y \approx \omega_0$, and $g(\omega)$ has a single peak around $\omega_0$ (see Figure \ref{fig-gplot}, right):
\begin{align}
    \Re \,g(\omega) =& 
    \frac{\gamma \omega ( \omega_y^2-\omega_x^2)}{
    \left(\left(\omega_x^2-\omega^2\right)^2+\left(\gamma\omega\right)^2\right)
    \left(\left(\omega_y^2-\omega^2\right)^2+\left(\gamma\omega\right)^2\right)
    }
    \label{reg}
    \\
    \approx&
    \frac{\gamma \omega ( \omega_y^2-\omega_x^2)}{
    \left((\omega_0^2-\omega^2)^2+(\gamma\omega)^2\right)^2
    }
    \\
    \approx& 
    \frac{\gamma \omega_0 ( \omega_y^2-\omega_x^2)}{
    \left((\omega_0^2-\omega^2)^2+(\gamma\omega_0)^2\right)^2
    }
    \\
    \approx&
    \frac{\gamma( \omega_y-\omega_x)}{8\omega_0^2}
    \frac{1}{
    \left((\omega-\omega_0)^2+\gamma^2/4\right)^2
    }
\end{align}
Using the limit of the Dirac delta function:
\begin{align}
    \frac{1}{\left(x^2+\varepsilon^2\right)^2} \rightarrow \frac{\pi}{2\varepsilon^3} \delta(x) \qquad\text{when}\qquad
    \varepsilon\to 0
\end{align}
we find that for small $\gamma$:
\begin{align}
    \Re\,g(\omega) \approx 
    \frac{\gamma( \omega_y-\omega_x)}{8\omega_0^2}\,
    \frac{4\pi}{\gamma^3} \delta(\omega-\omega_0)
\end{align}
Thus, the integral \eqref{XY-gphi} evaluates to:
\begin{align}    \label{xy-limit-1}
    \la XY\ra 
    =&
    -\frac{2\epsilon\omega_0}{\gamma}\,\frac{T}{\omega_0}\,\frac{\Omega(\omega_0)}{\omega_0^2-\Omega(\omega_0)^2}
    \\
    &\text{when}\qquad
    \epsilon \ll \frac{\gamma}{\omega_0} \ll 1
\end{align}
where we used $\omega_y-\omega_x=2\omega_0 \epsilon$.

\subsection{$\frac{\gamma}{\omega_0} \ll \epsilon \ll 1$}

In this limit, the width of the peaks is much smaller than the separation between $\omega_x$ and $\omega_y$ (Figure \ref{fig-gplot}, left).

First, approximate $g(\omega)$ around $\omega_x=\omega_0(1+\epsilon)$; starting from \eqref{reg}:
\begin{align}
    \Re\,g(\omega) \approx & 
    \frac{\gamma \omega_x ( \omega_y^2-\omega_x^2)}{
    \left(\left(\omega_x^2-\omega^2\right)^2+\left(\gamma\omega_x\right)^2\right)
    \left(\left(\omega_y^2-\omega_x^2\right)^2\right)
    }
    \\
    \approx& 
    \frac{\gamma }{8\omega_x^2\left(\omega_y-\omega_x\right)}
    \,\frac{1}{
    \left(\omega-\omega_x\right)^2+\gamma^2/4
    }
    \\
    \approx&
    \frac{\gamma }{8\omega_x^2\left(\omega_y-\omega_x\right)}
    \, \frac{2\pi}{\gamma}\delta(\omega-\omega_x),
\end{align}
where the last line uses the Dirac $\delta$ function approximation from \eqref{delta-limit}. Similarly, for the peak at $\omega_y$, we obtain:
\begin{align}
    \Re\,g(\omega) \approx&
    \frac{\pi }{4\omega_x^2\left(\omega_y-\omega_x\right)}
    \, \delta(\omega-\omega_x)
    \\
    &+
    \frac{\pi }{4\omega_y^2\left(\omega_y-\omega_x\right)}
    \, \delta(\omega-\omega_y).
\end{align}
Thus, the integral \eqref{XY-int} evaluates to:
\begin{align}
    \la XY \ra = &
    \frac{\pi \phi(\omega_x)}{4\omega_x^2\left(\omega_y-\omega_x\right)}
    +
    \frac{\pi \phi(\omega_y)}{4\omega_y^2\left(\omega_y-\omega_x\right)}
    \approx
    -\frac{\pi}{4\omega_0^3 \epsilon}
    \phi(\omega_0).\nonumber
\end{align}
Finally, we have
\begin{align}
    \la XY \ra = 
    -\frac{\gamma}{2\epsilon\omega_0}\,\frac{T}{\omega_0}\,\frac{\Omega(\omega_0)}{\omega_0^2-\Omega(\omega_0)^2} 
    \\
    \qquad\text{when}\qquad
    \frac{\gamma}{\omega_0} \ll \epsilon \ll 1.
    \label{xy-limit-2}
\end{align}

\subsection{$\epsilon \ll 1$ and $\gamma/\omega_0 \ll 1$}
In this section, we derive an approximate form of $\la XY \ra$ when $\gamma/\omega_0$ and $\epsilon$ are of similar orders of magnitude. Despite the coarse approximations employed, the final result remains in reasonable agreement with numerical evaluations.

As in the previous subsections, the integral \eqref{xyint} can be approximated by:
\begin{align}
\label{ftv}
    \la XY \ra = \Re\int_0^\infty \!\!\phi(\omega) g(\omega) \d \omega \approx \phi(\omega_0) \;\Re \int_0^\infty \!\!g(\omega) \d \omega
\end{align}  
since $g(\omega)$ exhibits significant variations primarily around $\omega_0$. However, the above integral cannot be evaluated analytically in its current form.

The denominator of $g(\omega)$ in \eqref{g-def} has four roots, denoted by $\omega_1, \omega_2, \omega_3, \omega_4$. Therefore, we express $g(\omega)$ as:
\begin{align}
    g(\omega) = \frac{\ii}{(\omega-\omega_1)(\omega-\omega_2)(\omega-\omega_3)(\omega-\omega_4)}.
\end{align}
The roots are given by:
\begin{equation}
    \label{omroots}
\begin{aligned}
    \omega_1 =& +\frac{\ii \gamma}{2}+\sqrt{\omega_x^2-\gamma^2/4},
    \quad \omega_3 = +\frac{\ii \gamma}{2}-\sqrt{\omega_x^2-\gamma^2/4},
    \\
    \omega_2 =& -\frac{\ii \gamma}{2}+\sqrt{\omega_y^2-\gamma^2/4},
    \quad
    \omega_4 = -\frac{\ii \gamma}{2}-\sqrt{\omega_y^2-\gamma^2/4}.
\end{aligned}
\end{equation}
All roots are complex and approach the real axis as $\gamma \to 0$. Only $\omega_1$ and $\omega_2$, which have positive real parts, significantly contribute to the rapid variations of $g(\omega)$.

We are primarily interested in the positive arguments of $g(\omega)$. Therefore, we consider the following approximation:
\begin{align}
    g_1(\omega) = \frac{\ii}{(\omega-\omega_1)(\omega-\omega_2)(\omega_0-\omega_3)(\omega_0-\omega_4)}
\end{align}
In the considered limit, $g(\omega) \approx g_1(\omega)$ for positive arguments, and $g_1(\omega) \approx 0$ for negative arguments. Hence, we can write:
\begin{align}
    \int_0^\infty  g(\omega) \d \omega \approx 
    \int_0^\infty  g_1(\omega) \d \omega
    \approx
    \int_{-\infty}^\infty  g_1(\omega) \d \omega.
\end{align}
The final integral can be evaluated analytically using the method of residues:
\begin{align}  
    \int_{-\infty}^\infty  g_1(\omega) \d \omega = 2\pi \ii \frac{\ii}{(\omega_1-\omega_2)(\omega_0-\omega_3)(\omega_0-\omega_4)}.
    \nonumber
\end{align}
Plugging in the roots \eqref{omroots}, we find: 
\begin{align}
    \epsilon \int_{-\infty}^\infty  g_1(\omega) \d \omega = - \frac{\pi}{4\omega_0^3}\frac{\epsilon}{\epsilon + \frac{\ii \gamma}{2\omega_0}} + o(\epsilon,\gamma).
\end{align}
Using \eqref{ftv}, we finally obtain:
\begin{align}
    \la XY \ra = -\frac{\epsilon\,\frac{\gamma}{2\omega_0}}{\epsilon^2+\left(\frac{\gamma}{2\omega_0}\right)^2} \frac{T}{\omega_0^2-\Omega(\omega_0)^2} \frac{\Omega(\omega_0)}{\omega_0},
    \label{xy-good}
\end{align}
which correctly reproduces the limits \eqref{xy-limit-1} and \eqref{xy-limit-2}.

\subsection{Moments of Phenomenologic Langevin Equation}
\label{app-pheno}

This section summarizes the findings from Ref.~\cite{matevosyan2024weak} and derives the $\langle XY \rangle$ moment of the stationary state distribution of the system. The Langevin equation \eqref{lang-pheno} with a harmonic anisotropic potential \eqref{V-aniso} can be rewritten as:
\begin{subequations}
    \label{OU}
\begin{align}
    &\dot{X_i}+ \Upsilon_{i j} X_j=\Gamma_i(t) ; \quad i=1, \ldots, 2d,
    \\
    &\left\langle\Gamma_i(t)\right\rangle=0, \quad
    \\
    &\left\langle\Gamma_i(t) \Gamma_j\left(t^{\prime}\right)\right\rangle=2 \rD_{i j} \delta\left(t-t^{\prime}\right), \quad \rD_{i j}=\rD_{j i},
\end{align}
\end{subequations}
where $d = 2$ and $(X_1, \dots, X_{2d}) = (\V, \R)$ represents the system's coordinates. Here, $(\Gamma_i)_{i=1}^{2d}$ is a white Gaussian noise with covariance matrix $2\rD$. The coefficients are defined as:
\begin{align}\label{FP-coefs}
    &\Upsilon = \begin{pmatrix}
        \gamma\; \rI & {\rm A}-\gamma\uo
        \\
        -\rI & 0
    \end{pmatrix},
    \qquad
    \rD=\begin{pmatrix} 
        \gamma T\,\rI & 0
        \\
        0 & 0
    \end{pmatrix},
    \\
    &\text{where}\quad 
    {\rm A}=\begin{pmatrix}
        \omega_0^2(1\!+\!\epsilon)^2 \!\!\!\!\!&0\\
        0&\!\!\!\!\!\omega_0^2(1\!+\!\epsilon)^2\!\!
    \end{pmatrix},
    \qquad
    \uo=\begin{pmatrix}
        0&-\Omega\\
        \Omega&0
    \end{pmatrix}.
\end{align}
The system of equations \eqref{OU} describes the Ornstein-Uhlenbeck process. The stationary state distribution is a Gaussian distribution with zero mean and covariance matrix $\sigma_\infty$:
\begin{eqnarray}
    \sigma_\infty = \begin{pmatrix}
        \Sigma_{\V\V} & \Sigma_{\R\V}\tr{}
        \\
        \Sigma_{\R\V} & \Sigma_{\R\R}
    \end{pmatrix}
\end{eqnarray}
which must satisfy the following equation:
\begin{eqnarray}\label{stationary-cond}
    \Upsilon\sigma_\infty+\sigma_\infty\Upsilon\tr{}=2 \rD.
\end{eqnarray} 
Solving this equation, we obtain:
\begin{align}
    \la XY\ra = -
    \tfrac{2 \gamma  T \Omega  \omega_0^2 \;\epsilon  \left(\gamma^2+\omega_0^2 \left(1+\epsilon^2\right)\right)}
    {\big(\gamma^2 \Omega^2+\omega_0^4 \left(1-\epsilon^2\right)^2\big) \left(\gamma^2 \left(\omega_0^2 \left(\epsilon^2+1\right)-\Omega^2\right)+4 \omega_0^4 \epsilon^2\right)}
    \label{XY-pheno-full}
\end{align}
In the limit $\gamma \ll \omega_0$ and $\epsilon \ll 1$, this simplifies to
\begin{align}
    \la XY\ra = \frac{\epsilon\,\frac{\gamma}{2\omega_0}}{\epsilon^2+
    \big(1-\frac{\Omega^2}{\omega_0^2}\big)
    \big(\frac{\gamma}{2\omega_0}\big)^2} 
    \frac{T \Omega}{\omega_0^3} 
\end{align}

\section{Free Brownian Motion Coupled to Rotating Thermal Bath}
\label{freefreeefree}

In this section, we investigate the transient distribution of a free Brownian particle coupled to a rotating thermal bath:
\begin{align}
    &\ddot{\R} = -\gamma \dot{\R} + \bxi.
\end{align}
where $\R=(X,\,Y,\,Z)$. For analytical calculations, we consider a linear angular velocity spectrum of bath oscillators, as in \eqref{linear}, so the autocorrelation of noise $\bxi$ is \eqref{C-1}. We assume that at time $t=0$ the particle starts from position $\R(0)=0$ and with velocity $\dot\R(0)=0$. The moments depending on only one coordinate are easy to calculate due to white noise on the diagonals of \eqref{noise-corr}:
\begin{align}
    &\la X^2(t)\ra = \la Y^2(t)\ra = \frac{q}{\gamma^2} t - \frac{q}{\gamma^3}\left(1-\ee^{-\gamma t}\right)\left(3-\ee^{-\gamma t}\right)
    \nonumber\\
    &\la V_x^2(t)\ra = \la V_y^2(t)\ra = \frac{q}{2\gamma}\left(1-\ee^{-2\gamma t}\right) 
    \nonumber\\
    &\la X(t) V_x(t)\ra = \la Y(t) V_y(t)\ra = 0
\end{align}
where $q=\frac{2\gamma T}{1-a^2}$ is the intensity of the white noise.
From the rotation symmetry of the system, we have:
\begin{align}
    \la X(t)Y(t)\ra = \la X(t)V_y(t)\ra = \la Y(t)V_x(t)\ra = 0.
\end{align}
Below we calculate the remaining moments $\la X(t)V_y(t)\ra = -\la Y(t)V_x(t)\ra$. We start with \eqref{XdX-int}:
\begin{align}
&\la X V_y\ra = \int_0^t\!\!\! \int_0^t\! K(t-u) K'(t-u') \la \xi_x(u) \xi_y(u')\ra \d u \d u'
\nonumber
\\
&~=-\frac{2\gamma T}{\pi}\frac{a}{1\!-\!a^2}\int_0^t\!\!\! \int_0^t \!K(t\!-\!u) K'(t\!-\!u') \calP \left(\tfrac{1}{u-u'}\right) \d u \d u'
\nonumber
\\
&~=\frac{2\gamma T}{\pi}\frac{a}{1-a^2}\int_0^t\!\!\! \int_0^t K(u) K'(u') \calP \left(\tfrac{1}{u-u'}\right) \d u \d u'
\end{align}
where Laplace kernel $K(t)$ is
\begin{align}
    K(t) = \frac{1}{\gamma}\left(1-\ee^{-\gamma t}\right), \quad K'(t) = \ee^{-\gamma t}
\end{align}
This consists of integrals of the form:
\begin{align}
    \label{app-F-def}
    F(t;\alpha,\beta) =& \int_0^t\int_0^t \ee^{-\alpha x}\ee^{-\beta y} \calP \left(\frac{1}{x-y}\right) \d x \d y
\end{align}
So, the moment $\la X(t)V_y(t)\ra$ can be expressed as:
\begin{align}\label{xvy-f-form}
    \la X(t)V_y(t)\ra =& §\frac{2T}{\pi}\frac{a}{1-a^2} 
    \left[ F(t;0,\gamma)-F(t;\gamma,\gamma) \right]
\end{align}
In order to evaluate \eqref{app-F-def}, recall the following elementary functions:
\begin{align}
    &\text{Exponential Integral:}\quad
    \Ei(x) = -\int_{-x}^\infty \frac{1}{t}\ee^{-t} \d t,
    \\
    &\text{Sinh Integral:}\quad
    \text{Shi}(x) = \int_0^x \frac{\sinh t}{t} \d t     \label{shi-def}
    \\
    &\hspace{22ex}=
    \frac{1}{2}\left(\Ei(x)-\Ei(-x)\right),
     \label{shi-def-2}
\end{align}
where $\frac{1}{t}$ in the integrals are interpreted as the Cauchy Principal Value, as defined in \eqref{Pval-def}. 
\begin{align}
    \int_0^t \ee^{-\alpha x} \calP \left(\frac{1}{x-y}\right) \d x =&
    \ee^{-\alpha y} \int_{-y}^{t-y} \ee^{-\alpha x} \calP \left(\frac{1}{x}\right) \d x
\end{align}
\comment{
First, assume $2y < t$:
\begin{align}
    \int_0^t \ee^{-\alpha x} \calP \left(\frac{1}{x-y}\right) \d x =&
    \ee^{-\alpha y} \left(\int_{-y}^{y}+ \int_{y}^{t-y}\right) \ee^{-\alpha x}\calP \left(\frac{1}{x}\right)\d x
    \\
    =& -2\ee^{-\alpha y} \text{Shi}(\alpha y) 
    +\left(E_1\right)
\end{align}

\begin{align}
    =&
    \begin{cases}
        \ee^{-\alpha y} \left(\int_{-y}^{y} \ee^{-\alpha x}\calP (\frac{1}{x})\d x
        +
        \int_{y}^{t-y} \ee^{-\alpha x}\calP (\frac{1}{x})\d x\right), & 2y < t
    \\
        \ee^{-\alpha y}\left(\int_{-y}^{y-t} \ee^{-\alpha x}\calP \frac{1}{x}\d x
        + \int_{y-t}^{t-y} \ee^{-\alpha x}\calP (\frac{1}{x})\d x\right), & 2y \geq t
    \end{cases}
    \\
    =&
    \ee^{-\alpha y} \left(\Ei(\alpha (y-t)) -\Ei(\alpha y) \right)
\end{align}
}
So for $F(t;\alpha,\beta)$ we have:
\begin{align}
    F(t;\alpha,\beta) =& \int_0^{t} \ee^{-\beta y} \ee^{-\alpha y} \left[\Ei(-\alpha (t-y)) -\Ei(\alpha y) \right] \d y \nonumber
    \\
    =&\,\ee^{-(\alpha+\beta) t}\int_0^t\ee^{(\alpha+\beta)y}\Ei(-\alpha y) \d y \label{first-int}
    \\
    &- \int_0^t \ee^{-(\beta+\alpha)y} \Ei(\alpha y) \d y
\end{align}
We apply integration by parts to each of these integral. The second integral reads:
\begin{align}
    &\int_\epsilon^t \ee^{-(\alpha+\beta)y} \Ei(\alpha y) \d y =
    \\
    &\quad=\frac{1}{\alpha\!+\!\beta}\Big[
    \!-\ee^{-(\alpha+\beta) y} \,\Ei(\alpha y) \Big|_\epsilon^t 
    +\!\int_\epsilon^t \frac{\ee^{-\beta y}}{y} \d y\,\Big]
\\    
&\quad= \frac{1}{\alpha+\beta}
    \Big(
        -\ee^{-(\alpha+\beta) t} \Ei(\alpha t) 
        +\Ei(-\beta t) 
\\
&\qquad\qquad\qquad+\Big[
          \ee^{-(\alpha+\beta) \epsilon}\Ei(\alpha \epsilon)
        - \Ei(-\beta \epsilon)
        \Big]
    \Big)
\end{align}
where we used $\Ei'(x)={\ee^x}/{x}$. Then, we take the limit $\epsilon \to 0^+$. For small negative arguments, $\Ei(-x) = \gamma_E + \ln x + O(x)$, where $\gamma_E$ is the Euler constant. Using \eqref{shi-def-2}, we change positive arguments to negative ones: $\Ei(\alpha \epsilon) = \Ei(-\alpha \epsilon) + 2\text{Shi}(\alpha \epsilon)$, and note that $\text{Shi}(0) = 0$. Thus, we have:
\begin{align}
&    \lim_{\epsilon\to 0^+}\left[
          \ee^{-(\alpha+\beta) \epsilon}(\Ei(-\alpha \epsilon) + 2\text{Shi}(\alpha \epsilon))
        - \Ei(-\beta\epsilon)
        \right] \nonumber\\
&\qquad  =\lim_{\epsilon\to 0^+}\left[\log(\alpha \epsilon) - \log(\beta\epsilon) +O(\epsilon)\right]
        \nonumber \\
&\qquad  =\log\left(\frac{\alpha}{\beta}\right)
\end{align}
We evaluate the first integral \eqref{first-int} in a similar way:
\begin{align}
&    \int_\epsilon^t \ee^{(\beta+\alpha)y} \Ei(-\alpha y) \d y =
\\
&\quad=
    \frac{1}{\alpha+\beta}\Big[
    \ee^{(\alpha+\beta) y} \Ei(-\alpha y) \Big|_\epsilon^t     -\frac{1}{\alpha+\beta}\int_\epsilon^t \frac{\ee^{\beta y}}{y} \d y
    \Big]
    \nonumber
\\
&\quad = \frac{1}{\alpha+\beta}
    \Big[
        \ee^{(\alpha+\beta)t} \Ei(-\alpha t) 
        -\Ei(\beta t) \nonumber\\
& \qquad\qquad\quad  -\Big(
          \ee^{(\alpha+\beta) \epsilon}\Ei(-\alpha \epsilon)
        - \Ei(\beta\epsilon)
        \Big)
    \Big]
\end{align}
The limit is also similar:
\begin{align}
&    \lim_{\epsilon\to 0^+}\left[\ee^{(\alpha+\beta) \epsilon}\Ei(-\alpha \epsilon)
        - \Ei(-\beta\epsilon) - 2\text{Shi}(\beta \epsilon)
        \right] =\nonumber\\
&        ~~~=\lim_{\epsilon\to 0^+}\left[\log(\alpha \epsilon) - \log(\beta\epsilon) +O(\epsilon)\right]
         = \log\left(\frac{\alpha}{\beta}\right)
\end{align}
Thus, we obtain:
\begin{align}
&    F(t;\alpha,\beta) = \frac{1}{\alpha+\beta}\Big[\Ei(-\alpha t)-\Ei(-\beta t)\\
&    +\ee^{-(\alpha+\beta) t} \left[\Ei(\alpha t) -\Ei(\beta t) \right]
 -\log(\alpha/\beta)(1+\ee^{-(\alpha+\beta) t})\Big]
 \nonumber
\end{align} 
We will also need this for $\alpha=0$ or $\beta=0$. Taking the corresponding limits:
\begin{align}
    F(t;0,\beta) =& \frac{1}{\beta}\Big[(\gamma + \log(\beta t))(1+\ee^{-\beta t})
    \\
    &\quad-\Ei(-\beta t) - \ee^{-\beta t} \Ei(\beta t)
    \Big],
    \\
    F(t;\alpha,0) =&\,-F(t,0,\alpha)
\end{align}
Turning back to the moment \eqref{xvy-f-form}, we have:
\begin{align}
&    \la X(t)V_y(t)\ra = \frac{2T}{\pi}\frac{a}{1-a^2} 
    \left[ F(t;0,\gamma)-F(t;\gamma,\gamma) \right]
 \nonumber   \\
& \qquad   =\frac{2T}{\pi}\frac{a}{1-a^2} \left[
        -\Ei(-t\gamma)-\ee^{-\gamma t}\Ei(\gamma t) \right.\nonumber\\
&  \qquad\qquad\qquad\qquad      \left.+(1+\ee^{-\gamma t})\left(\gamma_E + \log(\gamma t)\right)
    \right]
\end{align}
where $\gamma_E\approx0.58$ is the Euler constant. For small times, we get:
\begin{align}
    \la X(t)V_y(t)\ra =\frac{2T}{\pi}\frac{a}{1-a^2} \frac{\gamma t^2}{2}
\end{align}
and for large times it scales logarithmically:
\begin{align}
    \la X(t)V_y(t)\ra \sim \frac{2T}{\gamma\pi}\frac{a}{1-a^2} \log(t).
\end{align}

\end{document}